# Tracking the State and Behavior of People in Response to COVID-19 Through the Fusion of Multiple Longitudinal Data Streams

**Mohamed Amine Bouzaghrane***
ORCID: 0000-0003-1564-597
Department of Civil and Environmental Engineering
University of California, Berkeley
Email: amine.bouzaghrane@berkeley.edu

**Hassan Obeid**
Department of Civil and Environmental Engineering
University of California, Berkeley

**Drake Hayes**
Department of City and Regional Planning
University of California, Berkeley

**Minnie Chen**
Department of Civil and Environmental Engineering
University of California, Berkeley

**Meiqing Li**
Department of City and Regional Planning,
University of California, Berkeley

**Madeleine Parker**
Department of City and Regional Planning,
University of California, Berkeley

**Daniel A. Rodríguez**
Department of City and Regional Planning,
University of California, Berkeley

**Daniel G. Chatman**
Department of City and Regional Planning,
University of California, Berkeley

**Karen Trapenberg Frick**
Department of City and Regional Planning,
University of California, Berkeley

**Raja Sengupta**
Department of Civil and Environmental Engineering,
University of California, Berkeley

**Joan Walker***
Department of Civil and Environmental Engineering,
University of California, Berkeley,
Email: joanwalker@berkeley.edu

*Corresponding Author

**Abstract**
The changing nature of the COVID-19 pandemic has highlighted the importance of comprehensively considering its impacts and considering changes over time. Most COVID-19 related research addresses narrowly focused research questions and is therefore limited in addressing the complexities created by the interrelated impacts of the pandemic. Such research generally makes use of only one of either 1) actively collected data such as surveys, or 2) passively collected data from sources such as mobile phones or retail transactions. While a few studies make use of both actively and passively collected data, only one other study collects it longitudinally. Here we describe a rich panel dataset of active and passive data from U.S. residents collected between August 2020 and July 2021. Active data includes a repeated survey measuring travel behavior, compliance with COVID-19 mandates, physical health, economic well-being, vaccination status, and other factors. Passively collected data consists of all locations visited by study participants, taken from smartphone GPS data. We also closely tracked COVID-19 policies across counties of residence throughout the study period. Such a dataset allows important research questions to be answered; for example, to determine the factors underlying the heterogeneous behavioral responses to COVID-19 restrictions imposed by local governments. Better information about such responses is critical to our ability to understand the societal and economic impacts of this and future pandemics. The development of this data infrastructure can also help researchers explore new frontiers in behavioral science. The article explains how this approach fills gaps in COVID-19 related data collection; describes the study design and data collection procedures; presents key demographic characteristics of study participants; and shows how fusing different data streams helps uncover behavioral insights. The data are available to other researchers wishing to collaborate on future studies.

# 1. Introduction

The impacts of the COVID-19 pandemic are dynamic, apply at multiple spatial scales, and are unevenly distributed across individuals and space. Impacts such as loss of life, persistent health effects, changes in levels of employment and economic activity, and even changes in mobility, education, and civic and political engagement have been documented. This changing nature of the virus and its impacts, and our individual and societal responses to them, have highlighted the importance of being comprehensive in the outcomes considered and how they change over time. To date, however, most COVID-19 related research is cross-sectional and addresses a set of narrow research questions. This limits the ability to understand and address the complexities created by interrelated impacts surrounding the pandemic.

We designed a study to measure the impact of the COVID-19 pandemic on a wide range of aspects of human life and to track how these impacts change over time. Our data capture the impacts of the pandemic on several aspects of daily life as well as the longitudinal dynamics in human behavior, attitudes, and beliefs in response to the COVID-19 pandemic, using both passive and active data collection methods. This mix of active and passive data collection methods, gathered over several waves and covering a range of domains, stands in contrast to the majority of pandemic-related studies that have been more narrowly substantively focused and have made use of either active or passive data collection, but rarely both. Additionally, our data reflects the different scales at which behavior can be influenced; from the individual to the regional scale, making our data collection consistent with socio-ecologic approaches to understanding human behavior.

The combination of longitudinal active and passive data collection at different frequencies provides researchers with unique time-varying information addressing shortcomings of cross-sectional studies, and is useful in answering several questions surrounding the pandemic, using a diverse set of quantitative and qualitative methods. For example, as of the writing of this manuscript, our team has explored the impacts of COVID-19 on public transportation use (Parker et al. 2021) and is addressing topics including compliance with policy guidelines and mandate changes; residential relocation spurred by the pandemic; and persistent influences on telecommuting.

Our data's coverage of several domains is useful in addressing interdisciplinary questions and is the result of collaboration between transportation and non-transportation researchers. A holistic retrospective understanding of the impacts of the COVID-19 pandemic is fundamental to an effective management of the current state of the pandemic as well as future pandemics. More broadly, the development of this data infrastructure will help researchers explore frontiers in behavioral science useful in the management of large-scale disasters.

The purpose of this article is to present the data collection design and summarize characteristics of the sample collected. The rest of the article is organized as follows. Section 2 reviews previous data collection efforts surrounding COVID-19 and presents gaps addressed by our data. Section 3 describes the data infrastructure, study design, and adjustments made during the study period. Section 4 presents summary statistics about the data. Section 5 identifies conclusions and lessons gathered.

## 2. Literature

During the COVID-19 pandemic, a number of researchers examined the impacts of the virus on numerous aspects of human life, including mental and physical health, the economy, education, mobility, and the environment. Some researchers used actively collected data such as surveys, while others exploited passively collected data including smartphone use data, wearable technology data, and point of interest (POI) data. This section provides an overview of COVID-19 related research in the transportation literature and the data it used.

We mainly used Google Scholar to identify relevant COVID-19 research efforts. We used "COVID-19" and "coronavirus" as the primary keywords to initially identify relevant research, accompanied by other keywords indicating the research focus area. For transportation related studies, we focused on "travel behavior", "mode use", "driver behavior", "commute", "shared mobility", "e-commerce". We examined each relevant result to identify the research questions it addressed, whether it summarized the data collection effort it used, the content of the collected data, and whether the authors made their data available to other researchers or the public. In addition to results directly obtained from Google Scholar, we used the snowballing technique to identify other COVID-19 related research efforts.

Transportation researchers addressed a variety of questions. Researchers explored the impact of the pandemic on the number of trips (Abdullah et al. 2020; Beck and Hensher 2020; Fatmi 2020), mode use and mode shift (Abdullah et al. 2020; Beck and Hensher 2020; Bucsky 2020; de Haas et al. 2020; Shamshiripour et al. 2020; Eisenmann et al. 2021; Shakibaei et al. 2021), trip purpose (Abdullah et al. 2020; Beck and Hensher 2020; de Haas et al. 2020; Parady et al. 2020), distance traveled (Abdullah et al. 2020; Fatmi 2020; Lee et al. 2020; Molloy et al. 2020), public transit and active transportation (Jenelius and Cebecauer 2020; Nikiforiadis et al. 2020; Pawar et al. 2020; Teixeira and Lopes 2020; Chang et al. 2021; Dong et al. 2021; Eisenmann et al. 2021; Hu and Chen 2021; Przybylowski et al. 2021), commuting behavior (Abdullah et al. 2020; Pawar et al. 2020; Matson et al. 2021; Shakibaei et al. 2021), time spent traveling (Borkowski et al. 2021), and driving behavior (Katrakazas et al. 2020).

Out of all the studies we reviewed, only Matson et al. (2021) and Molloy et al. (2020) used both active and passive data. Both of these studies focused on the impact of COVID-19 on general human mobility in the United States and Switzerland, with Matson et al. (2021) tracking a sample of U.S. individuals in Spring 2020 and Fall 2020, and Molloy et al. (2020) tracking a sample of Swiss individuals throughout the COVID-19 pandemic. All other reviewed studies have made use of either active or passive data collection methods, with 65% of studies relying on survey/questionnaire data.

Out of the studies using solely survey/questionnaire data collection methods Parady et al. 2020, Shakibaei et al. (2021), Shamshiripour et al. (2020), and Beck and Hensher (2020) collected longitudinal data over short periods of time, while Chauhan et al. (2021) collected survey responses across two longitudinal waves all throughout the COVID-19 pandemic.

While the majority of studies focused on the impacts of the COVID-19 pandemic on general travel behavior, some focused on narrower research questions. For example, Eisenmann et al. (2021) present changes in mode use throughout the COVID-19 lock-downs in Germany, Nikiforiadis et al. (2020) explore the impact of COVID-19 on bikeshare usage in Greece, and Przybylowski et al. (2021) investigate the impact of the pandemic on public transit use in Poland.

Studies also vary in their geographical coverage. Approximately 85% of the reviewed covered specific countries or narrower geographical regions. However, Abdullah et al. (2020), Fraiberger et al. (2020), and Katrakazas et al. (2020) collected data from multiple countries around the world.

Studies using passive data had access to large sample sizes, with Lee et al. (2020) using aggregate level mobility data from 100 million individuals across the United States. Out of the reviewed transportation articles, Chauhan et al. (2021), Hu and Chen (2021), Zheng et al. (2020), and Molloy et al. (2020) made their data available to other researchers, either readily or by request.

In addition to academic researchers, technology companies collecting mobility data from their users have published it during the pandemic to help researchers and public health experts understand how mobility has changed in response to policies aimed at controlling the spread of the virus. Apple published aggregate (on city, county, and country level) mobility data showing the change in routing requests by travel mode (Walking, Driving, and Transit) compared to the baseline on January 13th, 2020 (Apple 2020). Similarly, Google has published aggregate community mobility reports broken down by types of locations visited throughout the pandemic. Locations tracked include residences, parks, grocery and pharmacy stores, transit stations, retail and recreation, and workplaces (Google 2020). Similarly, Grandata formed a partnership with the UN Development Program to make their mobile user data from 12 countries available to researchers (UNDP Latin America and the Caribbean 2020). In Switzerland, Intervista has conducted a study tracking the mobility changes of approximately 2500 individuals through a smartphone application and made this data available to the public. More specifically, Intervista focused on tracking changes in distances traveled for different activity purposes, mode use, and trip purpose (Intervista 2021)

From this review, we learned the following:

- Two out of 27 studies use both active and large-scale passive data collection methods
- Three out of 27 studies are longitudinal throughout the duration of the COVID-19 pandemic
- Eighteen out of 27 studies have a geographical scope of one country or more
- Four out of 27 studies make the data available to other researchers either openly or by request

Additionally, the content of the reviewed survey based studies varied and can be summarized as follows:

- Four out of 17 studies asked about safety measures taken by participants during COVID-19
- All 17 studies asked about participants' travel behavior, including travel behavior related attitudes
- Two out of 17 studies asked about participants' household dynamics
- Four out of 17 studies asked about participants economic circumstances
- Two out of 17 studies asked about participants' physical health
- Five out of 17 studies asked about participants' mental health and psychological traits
- Nine out of 17 studies attitudinal views towards COVID-19 and its related restrictions
- All studies collected participants demographic information

Our data complements the reviewed data collection efforts by: 1) combining large scale passively collected data with a smaller subset of actively collected survey data, 2) designing a survey that covers broader aspect of participants life and behavior including personality traits, political views, and vaccination intention and status, 3) deploying multiple waves of the survey throughout the COVID-19 pandemic, 4) deploying the survey to participants across the U.S., and 5) making our collected data accessible to other researchers. We acknowledge, however, that by being broader than other studies, we might not be able to capture deeper information on any singular aspect of human life during the pandemic. Section 3 goes into detail about our data collection effort.

# 3. Study Design and Administration

## 3.1. Data Infrastructure

We used a combination of data collection methods to develop a database that enables a broad understanding of how COVID-19 has affected people's behavior (**Figure 1**). First, we included passively collected data by Similarweb Inc., a mobile audience analytics company with a recruited panel representative of U.S. smartphone users, which comprises point of interest (POI) visit information and smartphone app use over time, as well as basic user socio-demographic data from a large sample of U.S. individuals. Second, we designed a longitudinal survey to capture a broad snapshot of people's behavior, beliefs, and attitudes in response to COVID-19. Third, given the variety in public health measures enacted in response to the pandemic, we tracked COVID-19 related policies to understand how individuals complied with public health directives. We combined these three data sources to create a dataset that captures a wide spectrum of human activity throughout the COVID-19 pandemic. This dataset can be integrated geographically with other external datasets, such as election results, COVID-19 statistics, and healthcare system capacity, to address a wider range of questions.

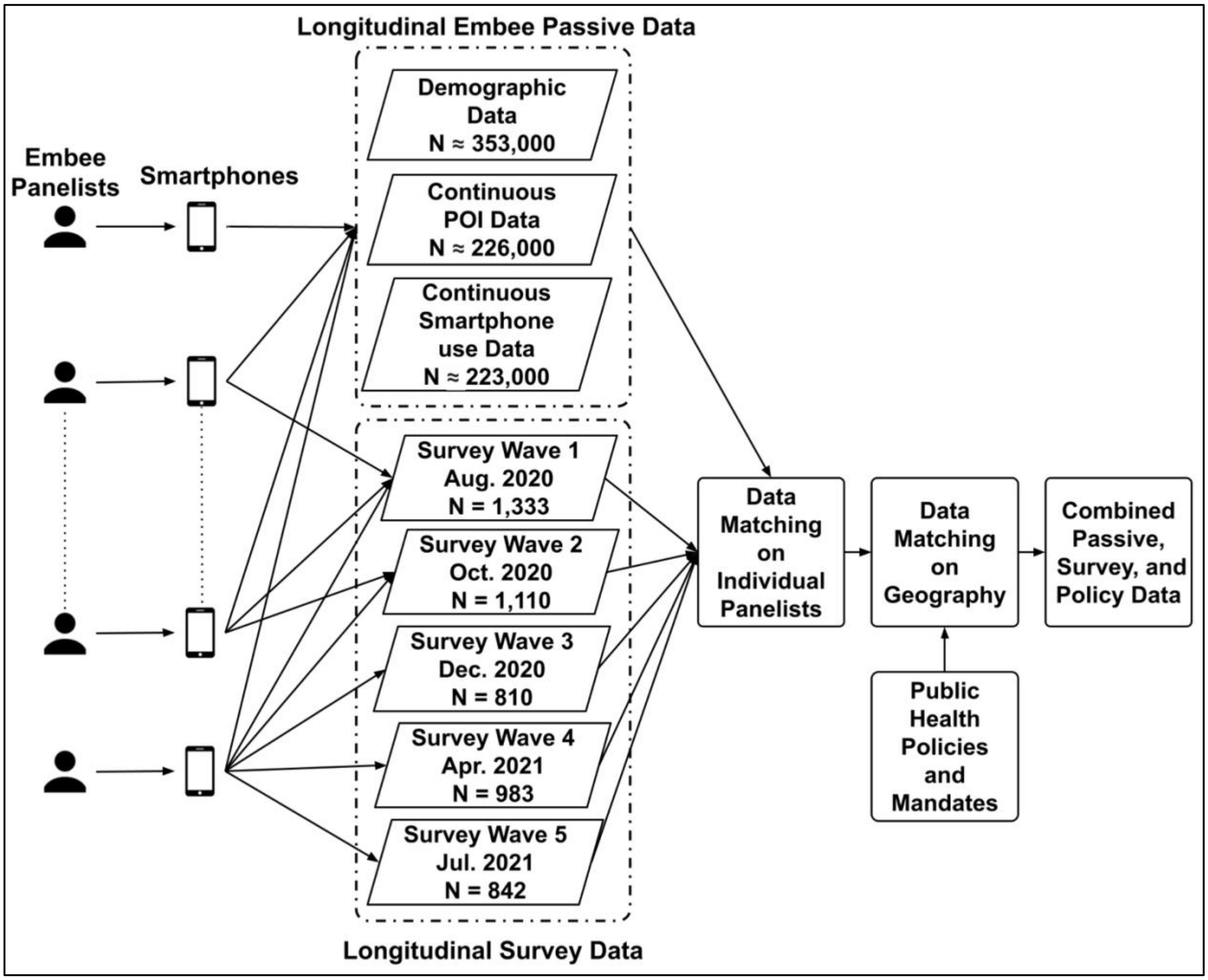

**Figure 1. Data infrastructure, N indicates the number of unique individuals in the dataset**

## 3.2. Study Timeline

An overview of the study timeline is shown in **Error! Reference source not found.**. We have passive data since January 2020. We started the initial survey data collection in early August 2020 and collected data in five total survey waves (August 2020, October 2020, December 2020, April 2021, July 2021). We also tracked stay-at-home/shelter-in-place and mask mandate policies across the states and counties of the study participants.

## 3.3. Similarweb Panel and Passive Data

Similarweb allows researchers to target recruited panelists based on criteria of interest. For example, one can target panelists based on their smartphone use behavior, socio-demographic characteristics, or geographic location. In addition to allowing researchers to collect survey data from its recruited panelists, Similarweb also partners with third-party partners to collect POI and smartphone app use data. The POI data includes information critical to inferring daily activities of panelists and understanding their daily travel behavior. This data are not continuously tracked GPS traces, but rather inferred individual check-ins at POIs using proprietary methods developed by Similarweb's third-party partners. For each individual check-in at a POI, the dataset includes information about the panelist's arrival and departure times, the category and brand of the location visited, the distance and time traveled to get to said location, the distance of the POI from the individual's identified home and work locations, as well as its zip code, city, and Metropolitan Statistical Area (MSA) name. Similarweb uses proprietary technology from a third-party provider to infer the location category from each of the POIs visited. The smartphone app use data provides a longitudinal description of smartphone use behavior for each of the recruited panelists. Each observation represents a smartphone activity and includes information about the app used as well as its duration of use.

## 3.4. Survey Design

We asked respondents about their economic well-being, mental health, physical health, personality type, political orientation, household dynamics, mobility behavior, living conditions, sheltering behaviors, preventative measures taken throughout the COVID-19 pandemic, and additional demographic information not collected by Similarweb. In designing our survey, we aimed to draw from validated survey questions in the literature. As such, our questionnaire contained a mixture of newly developed questions and validated questions. **Table *1*** summarizes the contents of the survey, including the source of any validated questions. The first survey wave included 11 sections, each of which focused on a distinct theme. The last section of the survey included open-ended questions providing study participants with the opportunity to share their thoughts on the pandemic, in addition to feedback about the survey. We altered the survey between waves to adapt to the changing pandemic context within the United States (e.g., relaxation and reimposition of restrictions, U.S. presidential elections, and vaccine availability). The length of the survey changed as we added and removed questions, but the main sections of the survey remained the same.

We required study participants to respond to all survey questions, with the exception of open ended questions. However, we included a "I prefer not to answer" option allowing respondents to opt out of responding to any question and to comply with the Institutional Review Board (IRB) guidelines. To reduce survey burden, we implemented conditional display logic whenever possible. To ensure high data quality, we implemented two attention check questions in different sections of the survey. We changed the placement of the attention check questions throughout the survey across survey waves to further reduce low-quality responses resulting from people remembering the survey flow from previous survey waves. The attention check questions specifically asked respondents to follow instructions and select specific choices. If a respondent failed to pass any of the attention check questions, they were dropped from the survey. The survey was approved by the University of California Committee for Protection of Human Subjects (CPHS). The first wave questionnaire is included in Appendix A.

**Table 1. Survey structure - *bolded questions are excerpted from validated surveys***

| Survey Section | Section Content | Source of Validated Questions | Changes in Subsequent Waves |
|---|---|---|---|
| **Safety Measures** | COVID-19 related safety measures, including handwashing and mask-wearing frequencies, the sizes of gatherings attended | | Included question about natural disasters |
| **Mobility** | Respondents' recent travel behavior, commuting/telecommuting behavior, ***main commute transportation modes***, ***vehicle ownership,*** attitudinal questions on use of ridersharing and public transit during COVID-19, recent purchases due to COVID-19 | U.S. National Household Travel Survey | |
| **Household Dynamics** | Respondents' household dynamics throughout the pandemic, including the number of individuals living in their household, the status of their relationship with their household members compared to before COVID-19, and whether the relationship with their household members affect their ability to spend time at home | | Added questions about changing primary residential location |
| **Economic Factors** | Respondents' change of employment since the beginning of COVID-19, changes in household income, ***financial stability,*** ability to sustain the economic and emotional impacts of the pandemic | U.S. Census, U.S. Federal Reserve | |
| **Political** | Respondents' pandemic news sources, knowledge and opinion of local pandemic restrictions, ***political affiliation,*** importance of religion, **willingness to get vaccinated**, and opinion about various pandemic related statements | PEW Research Center | Added questions about ***vaccination status*** and different ***political affiliation*** questions |
| **Personality** | ***Questions to measure personality*** | BFI-10 (Rammstedt and John 2007) | Removed for returning respondents |
| **Physical Health** | Questions about the respondents' physical health, their insurance status, level of worry about COVID-19, COVID-19 symptoms, COVID-19 testing status | | |
| **Psychological Factors** | Respondents' ability to be productive, feeling lack of companionship, ***anxiety and depression diagnosis*** | PHQ-4 (Kroenke et al. 2009) | |
| **Demographics** | ***General demographic information including details about the respondents living situation***, number of children in household, whether respondents provide care for a child or an elderly | U.S. Census | Added questions about disability, sexual orientation, and school attendance |
| **Open Ended** | Open ended questions asking about additional information on positive and negative aspects of the COVID-19 pandemic as well as feedback from respondents about the survey | | |

*Bolded questions are extracted from validated surveys

## 3.5 Sampling strategy

We developed a stratified sampling strategy to obtain a sample of panelists that was fairly geographically representative of the US population. Assuming a 10% response rate, we targeted the survey to approximately 14,500 panelists with a diverse mixture of metropolitan and rural counties across the US. We primarily focused on counties within 16 major MSAs across the United States, and selected a subset of counties within each to balance the number of panelists, area type and geographic distribution. To further balance our sample between rural and non-rural areas, we selected a set of rural counties across the US with the largest concentration of panelists. In total, 85% of targeted panelists are from metropolitan counties and 15% are from rural counties. **Figure 2** illustrates the geographical distribution of the targeted panelists from the first wave, where darker counties represent counties of targeted panelists.

The targeted sample over-represented people of color and lower income people compared to the US population. For example, only 50% of the targeted panelists were White/Caucasian in comparison to 72% of the US population. People with annual household income lower than $25,000 constituted 44% of the targeted sample, whereas only 20.2% of the US population fall into this income category. In subsequent survey waves, more specifically Wave 2 and Wave 4, due to a decrease in response rates, we augmented the target set with an additional randomly selected 10,000 and 5,000 panelists from across the United States, respectively. At the time of targeting, only 6,353 and 3,918 of the additionally sampled panelists were active.

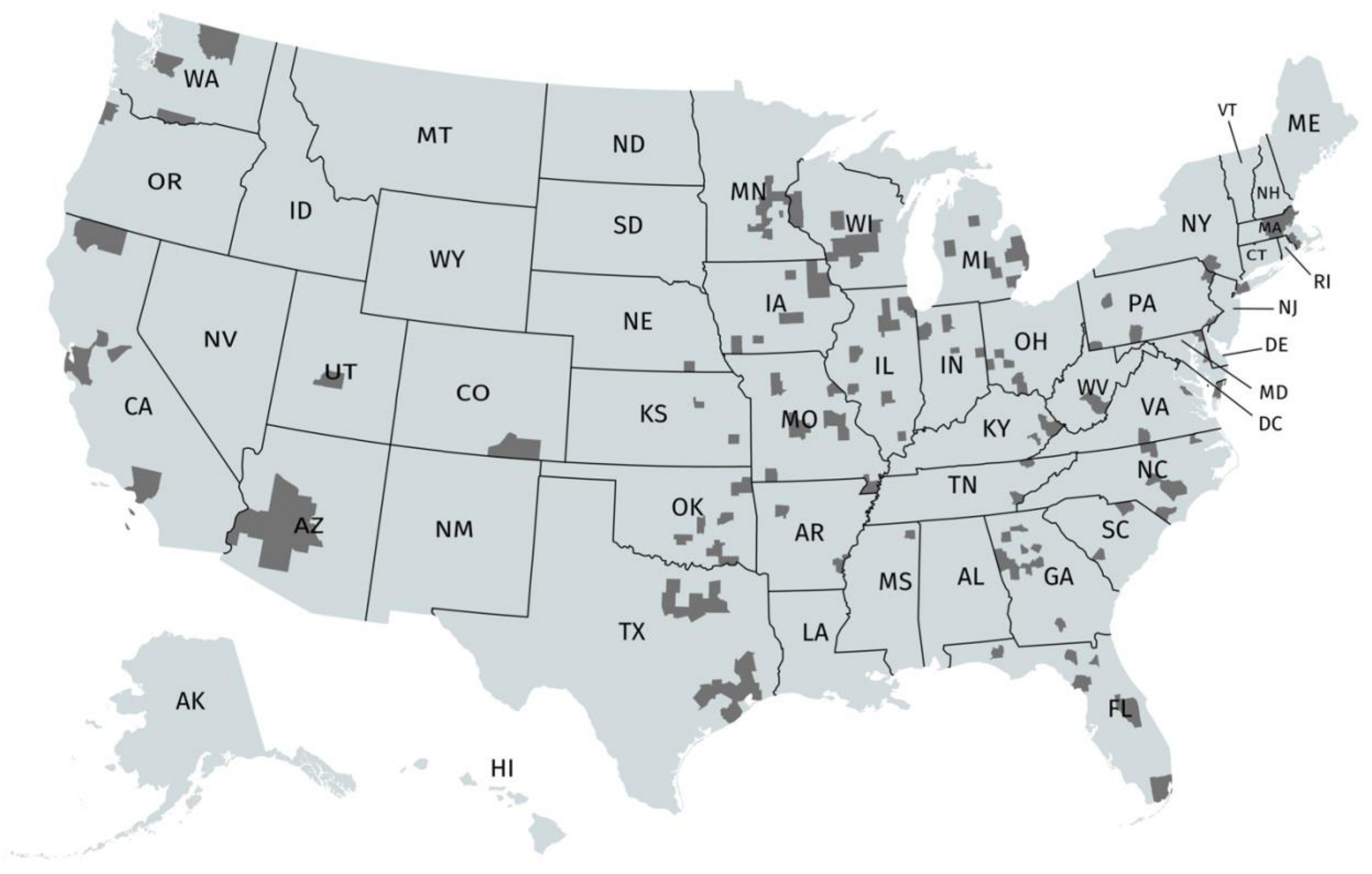

**Figure 2. Geographical distribution of targeted panelists**

## 3.6 Survey Pretesting

Due to the urgency of the study as well as existing budget restrictions, we conducted only informal pre-tests of the survey instrument with several graduate students from UC Berkeley and other individuals not affiliated with UC Berkeley. We used the Qualtrics survey platform to host and administer our questionnaire. We pre-tested the survey on Android smartphones, iPhones, and laptops. Following the feedback, we modified the survey by rearranging the order of several survey questions and added several survey cosmetic changes to improve the user experience. Additional comments from the survey pre-testers were mainly about the duration of the survey. Due to the large similarity between the different survey waves, we only tested subsequent surveys to ensure that the overall survey flow was not broken due to the addition or removal of survey questions. We also tested the survey with a small random sample of 20 Similarweb panelists prior to launching the first survey wave to all targeted panelists.

## 3.7 A-priori Survey Assessment

One of the determinants of response rates in any questionnaire is its complexity and length. We use the point scheme presented by Axhausen et al. (2015) to calculate the maximum survey burden score for each survey wave. The maximum survey burden is the sum of individual survey question burdens for the longest possible survey path a respondent could take.

**Table 2** presents fielding dates, the total number of questions, and the maximum calculated survey burden for each survey wave. The first survey wave has the largest a-priori survey burden because it has the largest number of questions. The mobility section accounts for the greatest share of the burden in the first three survey waves, while the political section is the most burdensome in the latest two waves. This is mainly due to the addition of several questions about vaccinations as vaccines became more available within the U.S. Within this section, the question asking respondents to identify the purposes for which each transportation mode was used has the largest survey burden. This question presents seven different transportation modes, for which up to six purposes could be selected. For a detailed breakdown of the response burden of each survey wave, please see Appendix B.

**Table 2. Respondent burden assessment using the point-based system proposed by Axhausen et al. (2015)**

| | **Wave 1 (Aug. 2020)** | **Wave 2 (Oct. 2020)** | **Wave 3 (Dec. 2020)** | **Wave 4 (Apr. 2021)** | **Wave 5 (Jul. 2021)** |
|---|---|---|---|---|---|
| **Deployment Period** | Aug. 3, 2020 - Sep. 12, 2020 | Sep. 26, 2020 - Nov. 2, 2020 | Dec. 4, 2020 - Jan. 3, 2021 | Mar. 26, 2021 - May 3, 2021 | Jun. 22, 2021 - Aug. 13, 2021 |
| **Number of Questions** | 76 | 75 | 67 | 72 | 73 |
| **Survey Burden (Points)** | 716 | 690 | 617 | 672 | 656 |
| **Largest Burden Section** | Mobility | Mobility | Mobility | Political | Political |

## 3.8 Study Participation

Given that all study participants had the Similarweb application installed on their smartphones, beyond answering the surveys, no additional effort was required from them to participate in our study. After each wave of the survey was made available, targeted panelists received notifications directly from the Similarweb smartphone application alerting them that they could answer the survey. Survey availability notifications stated that the survey was related to COVID-19, and provided recipients with an estimate of the time needed to complete it and the survey compensation amount.

Participants were asked to provide consent for each survey wave, after reviewing the purpose and scope of the data collection, a description of how the data was stored and protected, and agreeing to compensation for their participation. Respondents were provided with contact information of the principal investigator and were instructed to retain a copy of the informed consent form for their records. After starting the survey, respondents could leave at any time.

## 3.9 Study Participant Incentive and Participation Reminders

Our collaboration with Similarweb allowed us to access their panel, infrastructure, and data, free of charge. The most significant share of costs associated with this study were related to participant compensation. Only participants who answered all survey questions were compensated. Participants who failed the attention check questions were not allowed to complete the survey and were not compensated for their partial participation. Research shows that compensating study participants boosts participation rates (James and Bolstein 1992; Laguilles et al. 2011; Pedersen and Nielsen 2016) and makes it more likely to retain respondents in a longitudinal study (Yu et al. 2017).

Panelist retention proved to be a significant challenge throughout the study. As such, we increased financial compensation in subsequent waves to retain as many panelists as possible. The total compensation across the five completed survey waves added up to $33,200.

To further boost response rates, we sent reminders to targeted panelists starting from the second wave. Sending participation reminders can boost response rates up to 46% (Kongsved et al. 2007; Svensson et al. 2012; Van Mol 2017). We sent a single reminder in the second wave and daily reminders in the subsequent waves. We present detailed statistics on incentives, response rates, and completion times for each survey wave in the results section.

## 3.10 Policy Tracking

Throughout the COVID-19 pandemic, state and local governments took regulatory actions, known in the literature as Non-Pharmaceutical Interventions (NPIs), in an effort to decrease the transmission of the virus causing COVID-19. Various states and counties went into lockdown in March, implementing stay-at-home and shelter-in-place orders including curfews. With this action, restaurants, schools, and workplaces and some faith-based institutions closed in-person operations.

We began tracking these NPI's at the county level. We initially tracked 191 counties for stay-at-home orders, with the intention of tracking all 191 counties for mask mandates as well. However, given time constraints, we tracked 141 counties for mask mandates. Each of the counties that we did not collect mask mandate data for only had one respondent in our panel.

There were several occasions where county and state level orders co-occurred. In each circumstance, we checked for any state-level preemption of county orders. If that was the case, we recorded the current state regulations for that county. If there was no pre-emption, we recorded the stricter order of the two. For example, if the state did not have a stay-at-home order and the county did, we recorded a stay-at-home order being present in the county.

Over the course of collecting data, we initially tracked 191 counties, capturing every county represented in our panel. To streamline the tracking process, we examined numerous sources in an effort to find a county-level database with a rigorous data collection process. After not finding such a database, we made the decision to decrease the number of counties tracked because we did not have enough researcher hours to track and update information every 2 months. After the first survey wave in August 2020, we stopped tracking counties that had only 1 or 2 panelists, and continued to track 65 counties and 21 states. **Figure 3** shows the geographical distribution of the final 65 counties we tracked.

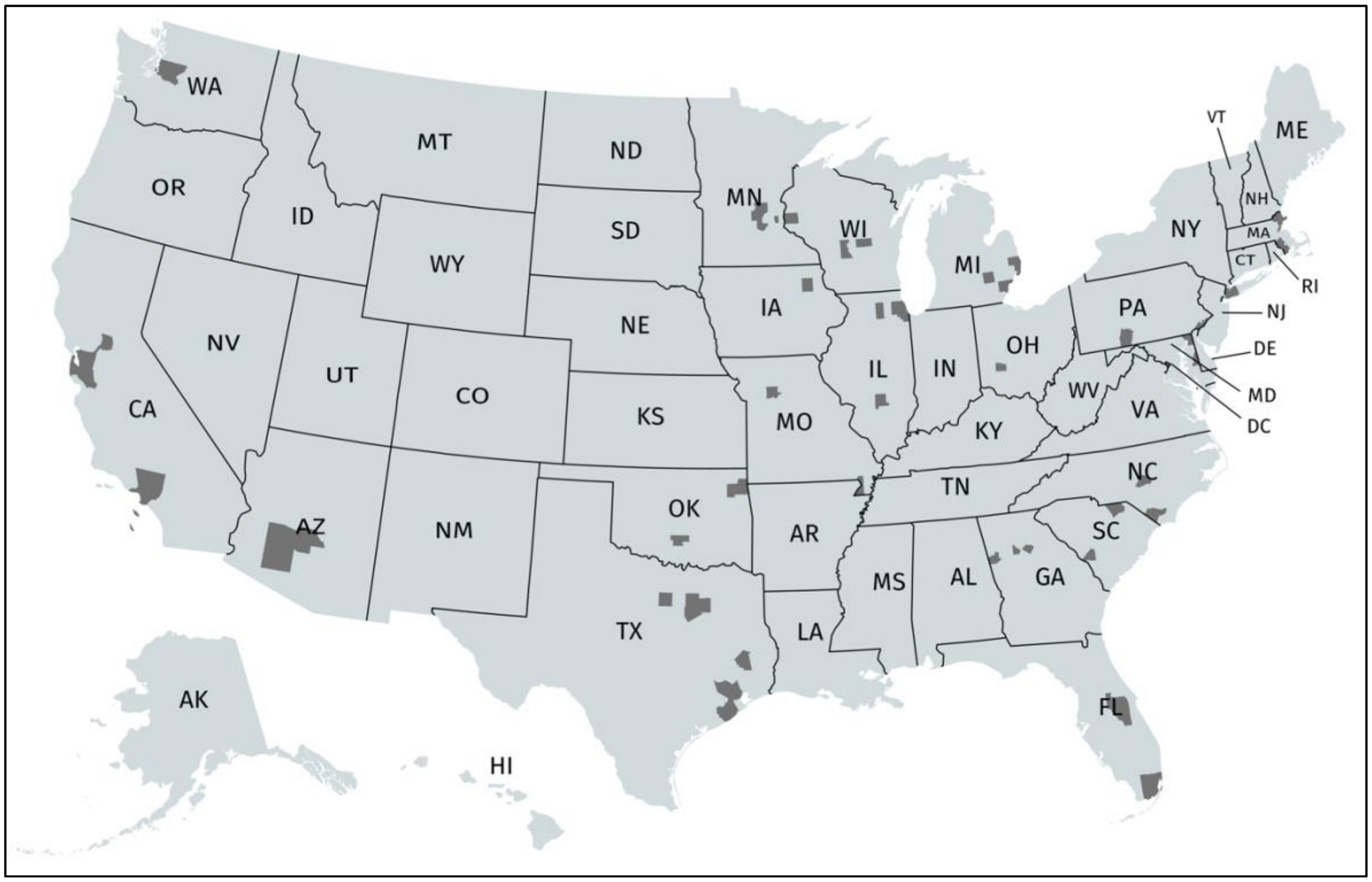

**Figure 3. Geographic policy tracking coverage**

# 4 Data and Results

## 4.1 Response Rates

Response rates for the surveys were calculated based on the web-based survey methodology and standards described by the American Association for Public Opinion Research (The American Association for Public Opinion Research 2016). We used the Response Rate 5 (RR5) model, which estimates the proportion of completed surveys out of all eligible respondents. We categorized panelist responses as follows: Completed surveys, accepted partial surveys, blank survey responses indicating consent, other surveys that were incomplete or failed the attention check, and unopened surveys. We defined "accepted" responses as those that answered all survey questions and did not fail any attention checks.

**Error! Reference source not found.** summarizes the number of responses, response rates, survey incentives, and completion times across the five survey waves. The response rate for the first wave was approximately 19%. Both the second and fourth survey waves targeted additional panelists, and had response rates of 14% and 18%, respectively. The third and fifth survey waves targeted only panelists who had responded to previous survey waves, and had significantly higher response rates (51% and 42% respectively). This indicates that individuals who responded to a previous survey wave are more likely to respond to subsequent survey waves when compared to newly targeted panelists. Panelist retention has proved to be a challenge in the study; out of the initial first wave respondents, 63% (847 panelists) completed the second wave, 42% (556 panelists) completed the first three waves, 25% (336 panelists) completed the first four waves, and 20% (262 panelists) completed all five survey waves.

The median survey completion time continuously declined through the duration of the study. This suggests that it became easier for panelists to complete the survey as they became more familiar with it.

**Table 4 – Number of Responses, Response Rates,Survey Retention, Survey Incentives, and Survey Completion Times**

| | | Wave 1 (Aug.2020) | Wave 2 (Oct.2020) | Wave 3 (Dec.2020) | Wave 4 (Apr.2021) | Wave 5 (Jul.2021) |
|---|---|---|---|---|---|---|
| **Targeted Sample (N)** | | 6968 | 7686 | 1586 | 5504 | 1962 |
| **Response Rate (%)*** | | 19 | 14 | 51 | 18 | 42 |
| **Sample Size (N)** | | 1333 | 1100 | 810 | 983 | 842 |
| **Share of Repeaters (%)** | | - | 85% | 100% | 62% | 100% |
| **Number of panelists with** | **One Previous Wave** | - | 847 | 254 | 53 | 277 |
| | **Two Previous Waves** | - | - | 556 | 218 | 86 |
| | **Three Previous Waves** | - | - | - | 336 | 217 |
| | **Four Previous Waves** | - | - | - | - | 262 |
| **Survey Incentive ($)** | | 4 | 5 | 5 | 8 | 11 |
| **Survey Completion Time** | **Median** | 20.0 | 17.0 | 16.0 | 17.0 | 14.5 |
| | **75$^{th}$ Percentile** | 28.0 | 24.0 | 24.0 | 25.0 | 20.5 |

**Figure 5** illustrates panelist retention across all deployed survey waves and shows the breakdown of participants in each survey wave by the most recent wave they participated in. The largest respondent drop occurred in the third survey wave. Additionally, the majority of respondents in each survey wave have participated in the preceding survey wave.

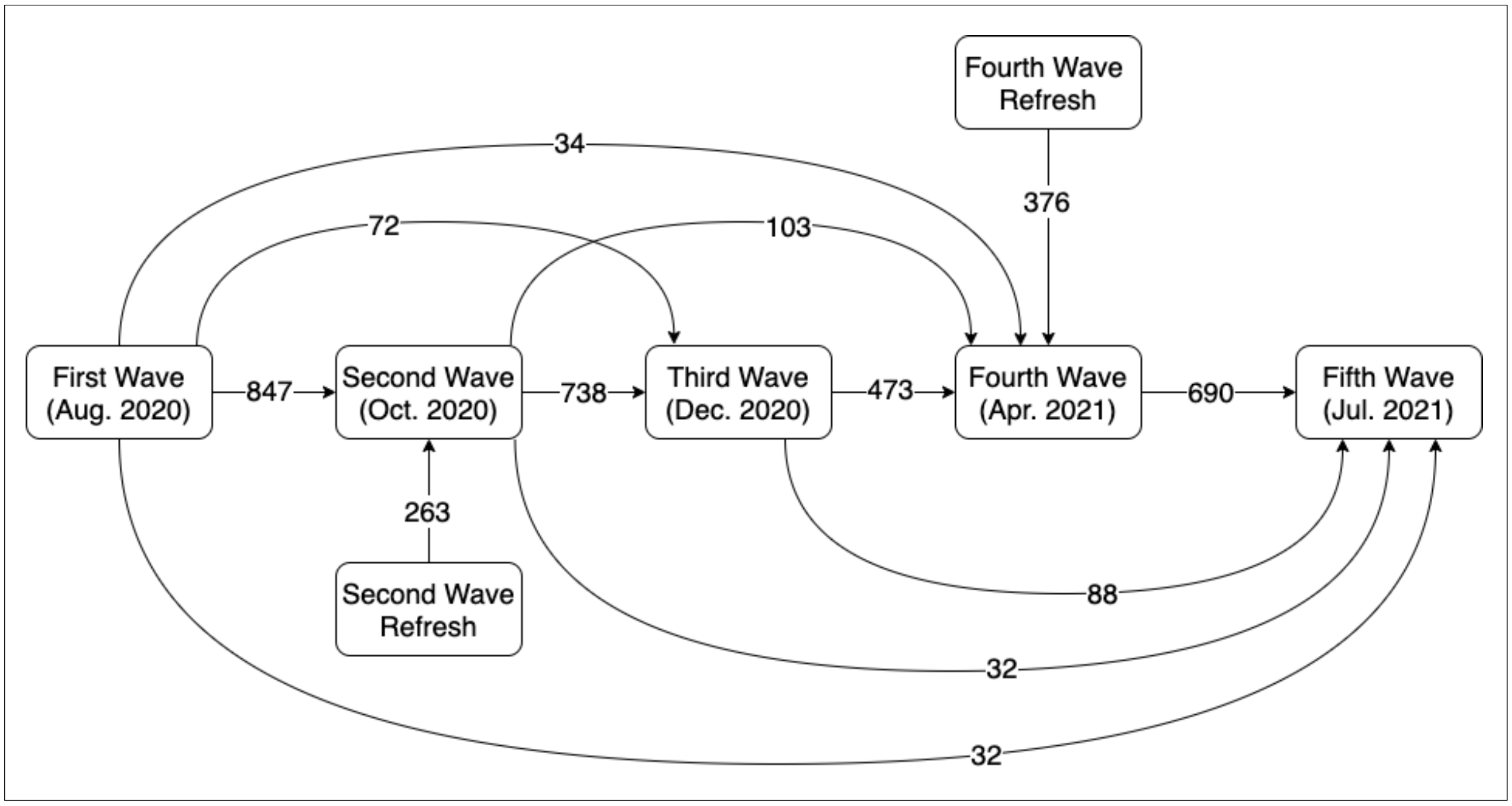

**Figure 4. Retention of survey respondents**

**Figure 5** illustrates the number of daily completed survey responses after the deployment of all five survey waves. The figure also shows when survey completion reminders were sent out to the respondents for each of the survey waves and whether survey compensation was increased. The number of daily responses is the highest immediately after each wave launch. We sent survey completion reminders starting from the second survey wave. Sending survey completion reminders in the second wave of the survey increased the daily completion rate immediately after sending the reminder. In the third wave of the survey, we sent daily reminders to the panelists, resulting in a slight increase in the number of daily responses immediately after the launch of the survey. The second and fourth wave graphs indicate that panelist pool refresh resulted in a significant jump in daily survey completions. We also increased survey compensation in the fourth and fifth survey waves. On one hand, it is unclear in the fifth wave whether the increase in daily survey completions was purely due to the pool refresh or the increase in survey compensation. On the other hand, daily completions from the fifth survey waves show that increasing survey compensation did not result in any significant increase in daily survey completions.

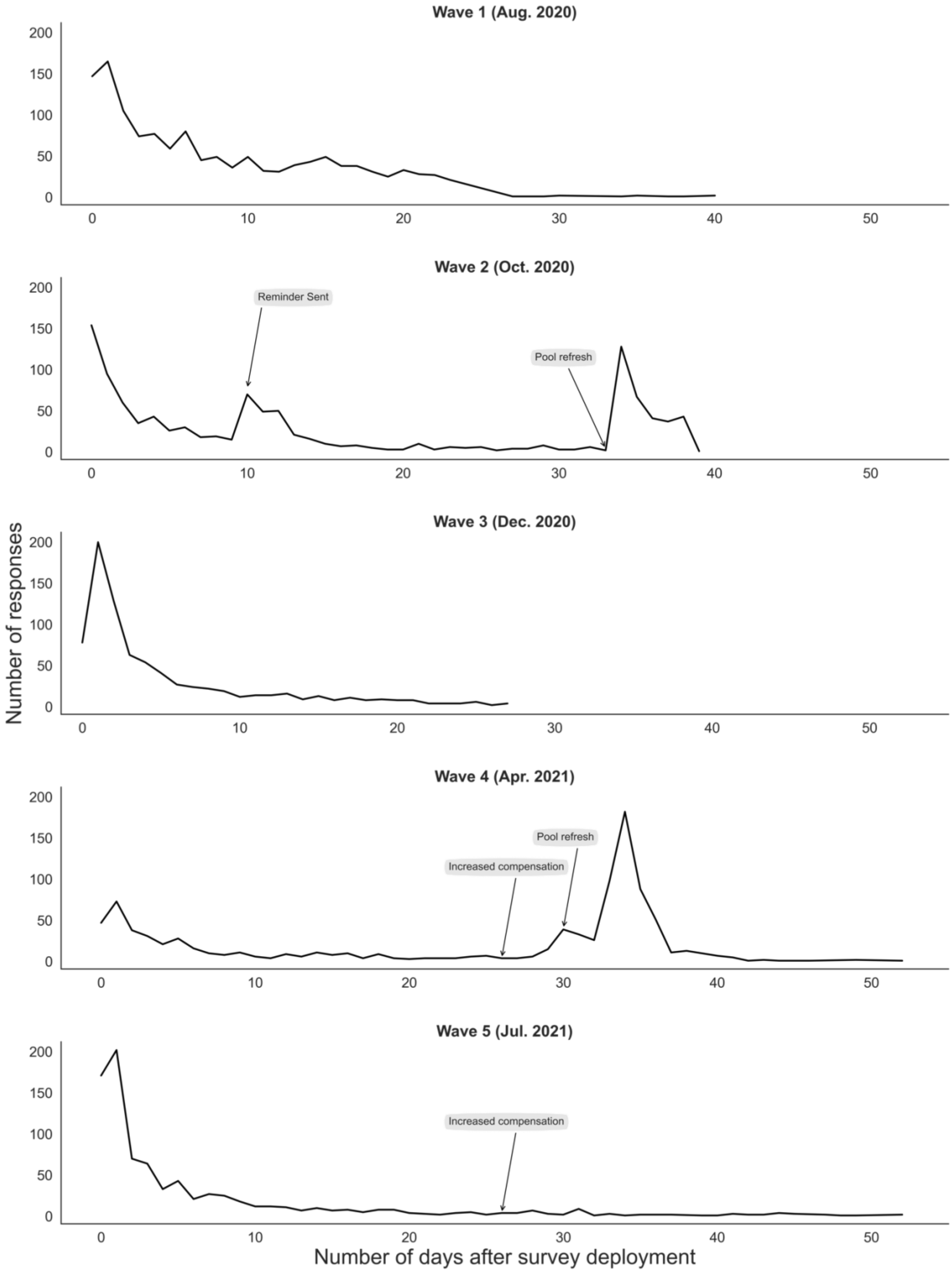

**Figure 5. Number of daily survey responses after deployment**

## 4.2 Demographic Summary

We present key demographic data for each survey wave in **Table 3** and **Table 4** and compare it to national statistics from the U.S. census where applicable. Our data oversamples women and undersamples men when compared to the US population; women respondents represent 55 to 60% of survey participants, compared to about 51% in the US population.

Our sample across all waves is heavily skewed towards young respondents, those between 25 and 54 years old. These respondents represent about 69 to 73% of all five survey waves, compared to 39% within the US population. Our sample also significantly undersamples individuals 19 years or younger (4% vs. 25%) and individuals 65 years and older (5% vs. 16%) across all survey waves compared to the U.S. population.

Our sample overrepresents households with low and medium income levels. For example, across all survey waves, between 59-65% of the respondents have an income of less than $50,000 compared to 40% in the U.S. population; and 10-11% of respondents have an income of $100,000 or higher, compared to about 31% in the U.S. population. It is possible that lower income individuals are more likely to be part of an online panel to earn income from survey incentives.

All survey waves undersample white Americans and oversample other groups. Caucasians represent approximately 72% of the U.S. population compared to approximately 52-56% of our respondents across waves. Black respondents represent about 13% of the U.S. population but comprise 18% of our respondents, across all five waves. Our sample underrepresents non-Hispanics/Latinos when compared to the U.S. population, with 75-77% of our sample identifying as a non-Hispanic/Latino compared to 82% of individuals in the U.S.

Our sample significantly underrepresents individuals with a high school education or less, who comprise 3-4% of our survey respondents, compared to 10% of individuals in the U.S. Those with a university or college degree represent between 43% and 46% across all survey waves compared to only 29% within the U.S. population. Finally, only 6% of our respondents hold a postgraduate degree, compared to 11% within the U.S. population.

Our study also undersamples smaller households. The share of 2-person or less households hovers around 40-45% in our data, compared to 62% within the U.S. population. On the other hand, 5-person or larger households comprise about 16-20% of our sample, compared to only 10% within the U.S. population.

The majority of our survey respondents have access to either 1 or 2 vehicles, representing between 70-73% of respondent households across all but the fourth survey wave. The share of panelists reporting working from home continuously decreased since the first survey wave, from 35% to approximately 20% in the fifth survey wave. This share is still significantly higher than that of the U.S. population pre-COVID19. Conversely, the number of individuals driving or carpooling to their workplaces increased from 50% in the first wave to 67% in the fifth wave, indicating the increasing propensity of individuals to commute to drive to work since the beginning of the COVID-19 pandemic, although still significantly less than the share of U.S. population pre-COVID19. Our sample is representative of public transit and active mode users (walking and biking). The share of these users has remained constant across all five survey waves and does not fully reflect the sinking share of transit use reported in the early stages of the pandemic by other researchers. Plausible factors behind changes in respondents' primary commute mode throughout the pandemic include reopenings of states across the nation and businesses requiring people to return to offices and places of work.

## Table 3. Demographic characteristics of study participants compared to the U.S. Census

| Category | Wave 1 (%) (Aug. 2020) | Wave 2 (%) (Oct. 2020) | Wave 3 (%) (Dec. 2020) | Wave 4 (%) (Apr. 2021) | Wave 5 (%) (Jul. 2021) | Targeted Counties (%) | Population (%) |
|---|---|---|---|---|---|---|---|
| **Gender** | | | | | | | |
| Male | 40.8 | 40.4 | 40.5 | 43.9 | 45.2 | 49.1 | 49.2 |
| Female | 59.2 | 59.6 | 59.5 | 56.1 | 54.8 | 50.9 | 50.8 |
| **Age** | | | | | | | |
| 19 years and under | 3.8 | 3.8 | 3.5 | 4.3 | 3.9 | 25.2 | 25.3 |
| 20 to 24 years | 9.7 | 9.0 | 8.1^* | 10.2 | 9.4 | 6.7 | 6.8 |
| 25 to 34 years | 23.7 | 22.4 | 23.0 | 23.7 | 24.1 | 14.6 | 13.9 |
| 35 to 44 years | 27.4 | 26.1 | 26.8 | 27.2 | 27.0 | 13.1 | 12.6 |
| 45 to 54 years | 20.2 | 21.2 | 22.0 | 20.4 | 21.3 | 13.3 | 13.0 |
| 55 to 59 years | 6.7^* | 7.3^* | 6.7^* | 5.7^* | 5.6^* | 6.6 | 6.7 |
| 60 to 64 years | 4.4 | 4.7^ | 4.7^ | 4.0 | 3.6 | 5.9 | 6.2 |
| 65 years and over | 4.3 | 5.5 | 5.5 | 4.6 | 5.2 | 14.5 | 15.6 |
| **Household Income** | | | | | | | |
| $0 - $24,999 | 35.1 | 31.8 | 30.4 | 30.0 | 27.8 | 17.7 | 19.3 |
| $25,000 - $49,999 | 30.1 | 32.5 | 30.8 | 28.7 | 29.1 | 19.2 | 21.2 |
| $50,000 - $99,999 | 24.5 | 25.4 | 27.3^* | 30.6^* | 31.1^* | 28.7 | 29.9 |
| $100,000 - $149,999 | 6.9 | 6.4 | 7.6 | 6.7 | 7.6 | 16.1 | 15.1 |
| $150,000 - $199,999 | 1.4 | 2.1 | 2.1 | 2.0 | 2.1 | 8.1 | 6.8 |
| $200,000 or more | 1.8 | 1.7 | 1.8 | 1.9 | 2.1 | 10.2 | 7.7 |
| **Race** | | | | | | | |
| Asian or Pacific Islander | 7.4* | 7.7^ | 8.4^ | 10.7^ | 11.0^ | 8.9 | 5.7 |
| Black/African American | 18.1 | 18.9 | 17.4 | 18.4 | 18.1 | 13.4 | 12.7 |
| Mixed Race | 6.5 | 6.4 | 5.7 | 7.1 | 6.8 | 3.6 | 3.3 |
| Native American/Alaskan Native | 2.5 | 2.8 | 3.1 | 1.9 | 1.9 | 0.6 | 0.9 |
| White/Caucasian | 54.5 | 54.7 | 56.5 | 52.2 | 53.3 | 65.6 | 72.5 |
| Other | 8.0^ | 7.0^ | 6.5^* | 6.3^* | 6.9^ | 7.8 | 4.9 |
| **Hispanic Status** | | | | | | | |
| Hispanic or Latino | 20.3* | 18.9* | 18.8* | 18.8* | 19.5* | 24.8 | 18.0 |
| Not Hispanic or Latino | 75.8^ | 77.9 | 77.5^ | 77.0^ | 77.3 | 75.2 | 82.0 |
| **Education Level** | | | | | | | |
| Less than High School | 3.5 | 3.1 | 2.8 | 3.8 | 2.9 | 13.1 | 10.1 |
| High School | 46.4 | 46.2 | 45.8 | 44.8 | 42.8 | 51.6 | 51.5 |
| University/College | 43.8 | 44.4 | 44.7 | 44.9 | 47.3 | 21.5 | 27.5 |
| Post-graduate Education | 6.1 | 6.2 | 6.5 | 6.4 | 7.1 | 13.7 | 11.0 |
| **Household Size** | | | | | | | |
| 1 | 15.1 | 16.3 | 14.2 | 15.4 | 14.8 | 27.0 | 28.0 |
| 2 | 25.2 | 25.8 | 26.3 | 26.3 | 26.3 | 31.9 | 34.0 |
| 3 | 20.8 | 21.5 | 24.1 | 20.4 | 23.1 | 16.2 | 15.6 |
| 4 | 18.3 | 16.2 | 16.8 | 19.6 | 18.7 | 14.0 | 13.0 |
| 5 | 10.6 | 11.2 | 9.6 | 8.4 | 7.7^* | 6.6 | 6.0 |
| 6+ | 9.8 | 8.4 | 8.4 | 9.2 | 8.4 | 2.6 | 2.3 |

*indicates a statistic representative of the U.S. population at the 5% level

^indicates a statistic representative of the population at the targeted counties at the 5% level

Table 4. Transportation-related descriptive statistics

| Category | Wave 1 (%) (Aug. 2020) | Wave 2 (%) (Oct. 2020) | Wave 3 (%) (Dec. 2020) | Wave 4 (%) (Apr. 2021) | Wave 5 (%) (Jul. 2021) | Targeted Counties (%) | Population (%) |
|---|---|---|---|---|---|---|---|
| **Household Vehicle Ownership** | | | | | | | |
| 0 | 12.4^ | 12.3^ | 11.7^ | 10.2^* | 8.9* | 11.4 | 8.6 |
| 1 | 38.2 | 39.9 | 37.0 | 36.7 | 39.0 | 32.9 | 32.7 |
| 2 | 32.5 | 33.1^ | 34.1^ | 21.0 | 34.5^* | 35.6 | 37.2 |
| 3 or more | 16.6 | 14.0 | 16.5 | 31.5 | 17.2 | 19.9 | 21.4 |
| **Primary Commute Mode** | | | | | | | |
| Not Applicable | 34.3 | 24.3 | 28.0 | 21.2 | 19.5 | 5.3 | 5.2 |
| Car † | 50.8 | 61.0 | 58.2 | 65.6 | 67.7 | 80.4 | 85.3 |
| Carsharing | 0.8 | 0.6 | 0.9 | 0.3 | 0.0 | N/A | N/A |
| Ridehail or Taxi | 2.0 | 1.2 | 1.7 | 2.3 | 1.9 | 0.1 | 0.2 |
| Transit | 6.0* | 6.1* | 6.7* | 5.5* | 5.5* | 9.5 | 5.0 |
| Bicycle | 0.6^* | 0.9^* | 0.4^* | 0.7^* | 0.4^* | 0.6 | 0.5 |
| Walking | 2.2^* | 3.3^* | 3.1^* | 2.5^* | 2.7^* | 3.0 | 2.7 |
| Other | 1.6^ | 1.4^* | 0.5^* | 0.3 | 0.6^* | 1.1 | 1.1 |

*indicates a statistic representative of the U.S. population at the 5% level

^indicates a statistic representative of the population at the targeted counties at the 5% level

† Drive Alone and Carpooling are aggregated into one category to compare U.S. census statistics to our survey data

## 4.3 Incomplete Surveys

As with all surveys, some respondents will not stay through to completion. Survey incompletion could be due to several reasons, including survey length, nature of the survey questions, relevancy of the questions to respondent, survey incentive amounts, or technical difficulties. Galesic (2006) found that survey incompletion is associated with higher experienced survey burden and overall lower interest in participation. More specifically, as mentioned in Section 4.1, given the overall conditions within the United States throughout the COVID-19 pandemic, the research team suspected that the University of California brand could impact our response rate and data quality. Fang et al. (2012) show that a research project's sponsoring corporation reputation can have a significant impact on people's willingness to participate in a web-based survey. When survey incompleteness is not random, it could prevent results from being generalizable to the population.

For our objectives, any survey respondent that did not complete the survey is considered to have abandoned the survey. **Table 5** summarizes the share of complete and incomplete survey responses. Incomplete responses can further be categorized into those due to respondents dropping out of the survey and those due to the respondent's failure to properly answer the attention check questions. These statistics are computed relative to the number of participants who opened the survey, as opposed to all individuals initially targeted. The share of survey completes increased between the first and fifth wave from 66% to 90%. The share of survey incompletion was 10% in the first survey wave, consistent with findings from Hoerger (2010) and decreased significantly to 3% in the fifth survey wave. This decrease can possibly be explained by the commitment and interest of panelists staying on the panel throughout the study period. Similarly, and possibly due to similar reasons, the share of panelists failing the attention check question decreased by upwards of 70% from the first survey waves (from 24% to 7%). Lower survey incompletion rates in later survey waves suggests that retained panelists are panelists with greater interest and attention.

**Table 5. Share of participants with complete, incomplete, and failed attention check responses**

| | Wave 1 (Aug. 2020) | Wave 2 (Oct. 2020) | Wave 3 (Dec. 2020) | Wave 4 (Apr. 2021) | Wave 5 (Jul. 2021) |
|---|---|---|---|---|---|
| **Complete (%)** | 66 | 79 | 85 | 73 | 90 |
| **Incomplete (%)** | 10 | 4 | 2 | 9 | 3 |
| **Failed Attention Check (%)** | 24 | 17 | 14 | 18 | 7 |

**Table 6** illustrates the share of study participants exiting the survey at key points through the survey: before the mobility section, at the mobility section, and after the mobility section. The table shows that a significant share of survey incompletion occurred immediately at the beginning of the survey in the first survey wave. This initially high dropout rate was also significant in the fourth wave as we targeted additional panelists. The dropout rate was also significant at the mobility section across all five survey waves, higher than that of any other section throughout the survey. As mentioned earlier, several questions in this section had large burden scores. For example, a series of questions in this section ask about the respondents' pandemic transportation behavior (e.g., frequency and usage purpose of several transportation modes). These questions were designed in repetitive matrix format with several options to select. Additionally, given that all study participants answered surveys on their smartphones, these questions might not have been easily accessible to all participants given different screen sizes of different smartphone models.

**Table 6. Share of survey dropout at key survey sections**

| | Wave 1 (Aug.2020) | Wave 2 (Oct.2020) | Wave 3 (Dec.2020) | Wave 4 (Apr.2021) | Wave 5 (Jul.2021) |
|---|---|---|---|---|---|
| **Incomplete (%)** | 14 | 5 | 1.7 | 10 | 3 |
| **Before Mobility Section (%)** | 4 | 1 | 0 | 3 | 0 |
| **At Mobility Section (%)** | 5 | 3 | 0.7 | 3 | 2 |
| **After Mobility Section (%)** | 3 | 1 | 1 | 4 | 1 |

## 4.4 Study Participation Models

We estimated several binary logit models to help describe the factors associated with the following outcomes:

- Whether a panelist opened the survey
- Whether a panelist did not complete the survey
- Whether a panelist failed the attention check questions
- Whether a panelist from the first wave participated in the second wave of the survey

For comparability we maintained the same model specification across all four models. **Table 7** presents specification and results. Age is coded as the mid-point of the age categories described in **Table *3***. Household size was self-reported by respondents. The remaining variables in the model specification are all binary variables. For example, the variable "More Than 30 Days on Panel" indicates whether a panelist has been on the Similarweb online panel for more than 30 days at the time of the deployment of the first survey wave with a "1" value, and "0" otherwise.

The model results are mixed. Duration on the panel is the variable most highly associated with whether a targeted panelist opens the survey, with individuals on the panel for longer than 30 days being significantly less likely to open the survey. Surprisingly, individuals with higher household income levels are more likely to open the survey. Full-time workers are less likely to open the survey, suggesting a possible lack of available time to participate.

Additionally, non-White, Hispanic, less-educated, and lower income participants are more likely to fail the survey's attention check questions included in the survey. Full-time workers are also more likely to fail attention check questions. This could suggest that these individuals were trying to complete the survey faster and could have missed the instructions shown in the attention check questions.

The results also indicate that non-white participants are more likely to not complete the survey. This could suggest that our survey design and language might not be as accessible or interesting to people of color. When investigating the factors associated with respondents returning to participate in the second survey wave, we find that higher income individuals were more likely to participate.

**Table 7. Model results summary**

| | Open Survey | Not Complete Survey | Fail Attention Check | Participate in Second Wave |
|---|---|---|---|---|
| **Intercept** | -0.898*** | -2.757*** | -1.126*** | 0.692** |
| | (0.147) | (0.436) | (0.335) | (0.35) |
| **Age** | 0.006*** | 0.008 | -0.009* | 0.004 |
| | (0.002) | (0.006) | (0.005) | (0.005) |
| **Male** | -0.257*** | -0.134 | 0.428*** | -0.016 |
| | (0.051) | (0.157) | (0.120) | (0.127) |
| **Non-white** | -0.073 | 0.638*** | 0.274** | -0.036 |
| | (0.051) | (0.159) | (0.121) | (0.128) |
| **Hispanic** | 0.110* | -0.118 | 0.366*** | -0.124 |
| | (0.062) | (0.187) | (0.137) | (0.153) |
| **Household Size** | -0.051*** | 0.076* | 0.014 | -0.095*** |
| | (0.014) | (0.042) | (0.033) | (0.037) |
| **Education Level: University/College** | 0.066 | -0.076 | -0.487*** | -0.150 |
| | (0.054) | (0.165) | (0.132) | (0.132) |
| **Education Level: Postgraduate** | 0.072 | -0.193 | -0.240 | -0.156 |
| | (0.112) | (0.361) | (0.270) | (0.275) |
| **Income: $25,000 - $49,999** | 0.149** | -0.107 | -0.620*** | 0.278* |
| | (0.065) | (0.191) | (0.153) | (0.157) |
| **Income: $50,000 - $99,999** | 0.320*** | -0.259 | -0.595*** | 0.291* |
| | (0.072) | (0.221) | (0.171) | (0.175) |
| **Income: $100,000 - $149,999** | 0.522*** | -0.311 | -0.672** | 0.194 |
| | (0.115) | (0.370) | (0.289) | (0.266) |
| **Income: $150,000 - $199,999** | 0.663*** | -0.593 | -0.205 | 1.571** |
| | (0.218) | (0.755) | (0.463) | (0.765) |
| **Income: $200,000 or more** | 0.630*** | -0.100 | -0.245 | 0.458 |
| | (0.193) | (0.562) | (0.428) | (0.497) |
| **Full Time Worker** | -0.115** | -0.070 | 0.341*** | -0.133 |
| | (0.056) | (0.170) | (0.130) | (0.136) |
| **More Than 30 Days on Panel** | -1.157*** | -0.038 | 0.067 | 0.392* |
| | (0.098) | (0.264) | (0.207) | (0.204) |
| **Observations** | 14,581 | 1,871 | 1,871 | 1,330 |
| $\rho^2$ | 0.019 | 0.019 | 0.042 | 0.014 |

**∗p<0.1; ∗∗p<0.05; ∗∗∗p<0.01**

## 4.5 Tracking COVID-related Policies

After March's rush of stay-at-home and shelter-in-place orders (**Figure 6**), governments began mandating the use of masks and face coverings in public places. Initially, public health experts discouraged the use of face masks to ensure there were enough masks and personal protective equipment for frontline healthcare workers (Panetta 2020). As COVID-19 continued to spread, and research supporting the effectiveness of masks emerged, many regions implemented face covering mandates (Molteni and Rogers 2020). Most orders started between April and July 2020 and only expired in a handful of the counties we tracked.

As can be seen from **Figure 6** and **Figure 7**, 93% of our initially tracked counties had stay-at-home orders, along with 68% of our initially tracked counties. Research analyzing mask wearing mandates in the United States has discovered the best indicator for a state not having a mandate is a state having a Republican governor (Adolph et al. 2020). Our unique county-level dataset will allow us to conduct similar research at the county-level, uncovering why orders were implemented and whether they were followed.

As of January 2021, our NPI tracking shows a patchwork of policies across the United States. Certain NPIs only apply if a county or region reaches a certain case or intensive care unit capacity threshold.

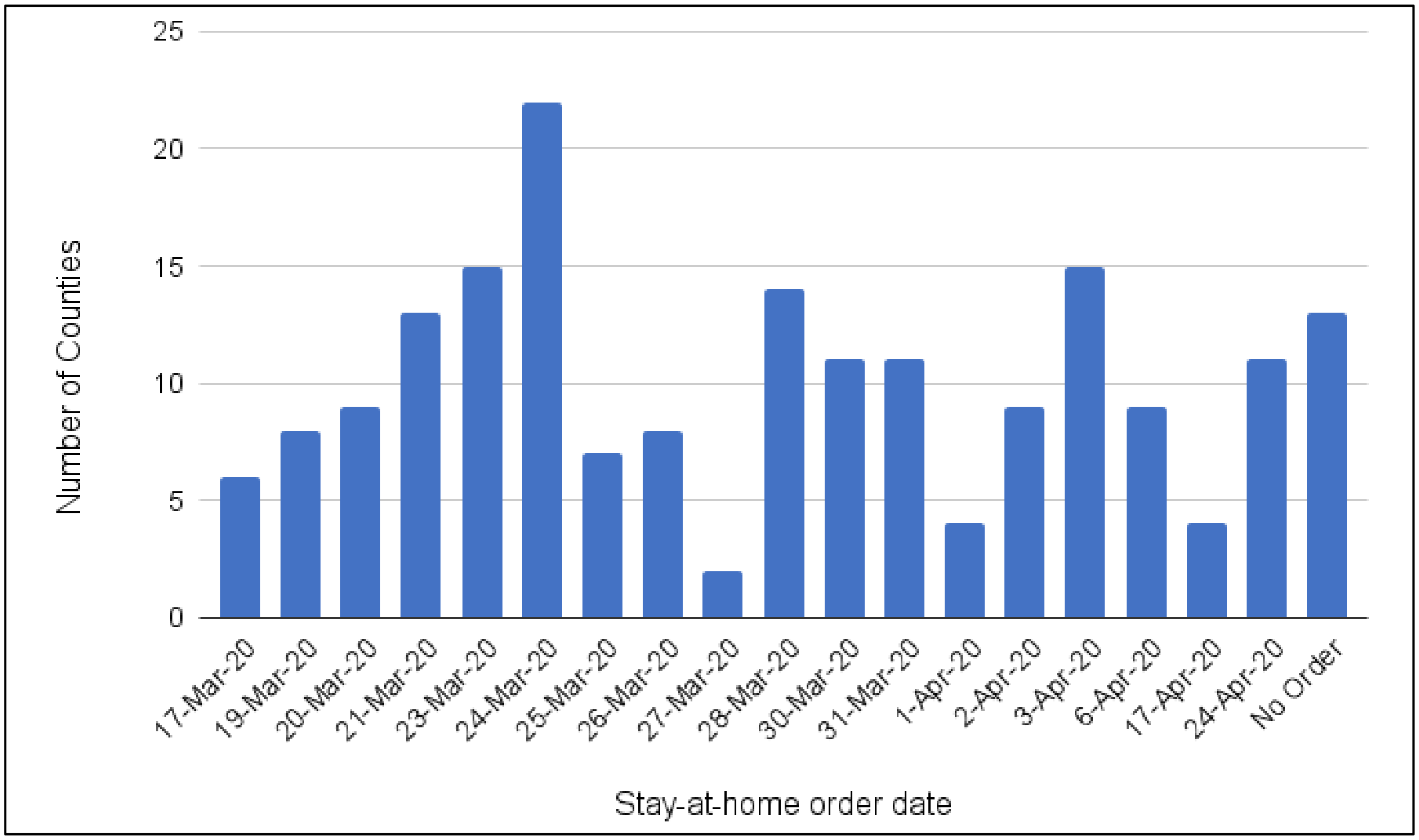

**Figure 6. Distribution of county-level stay at home order start dates**

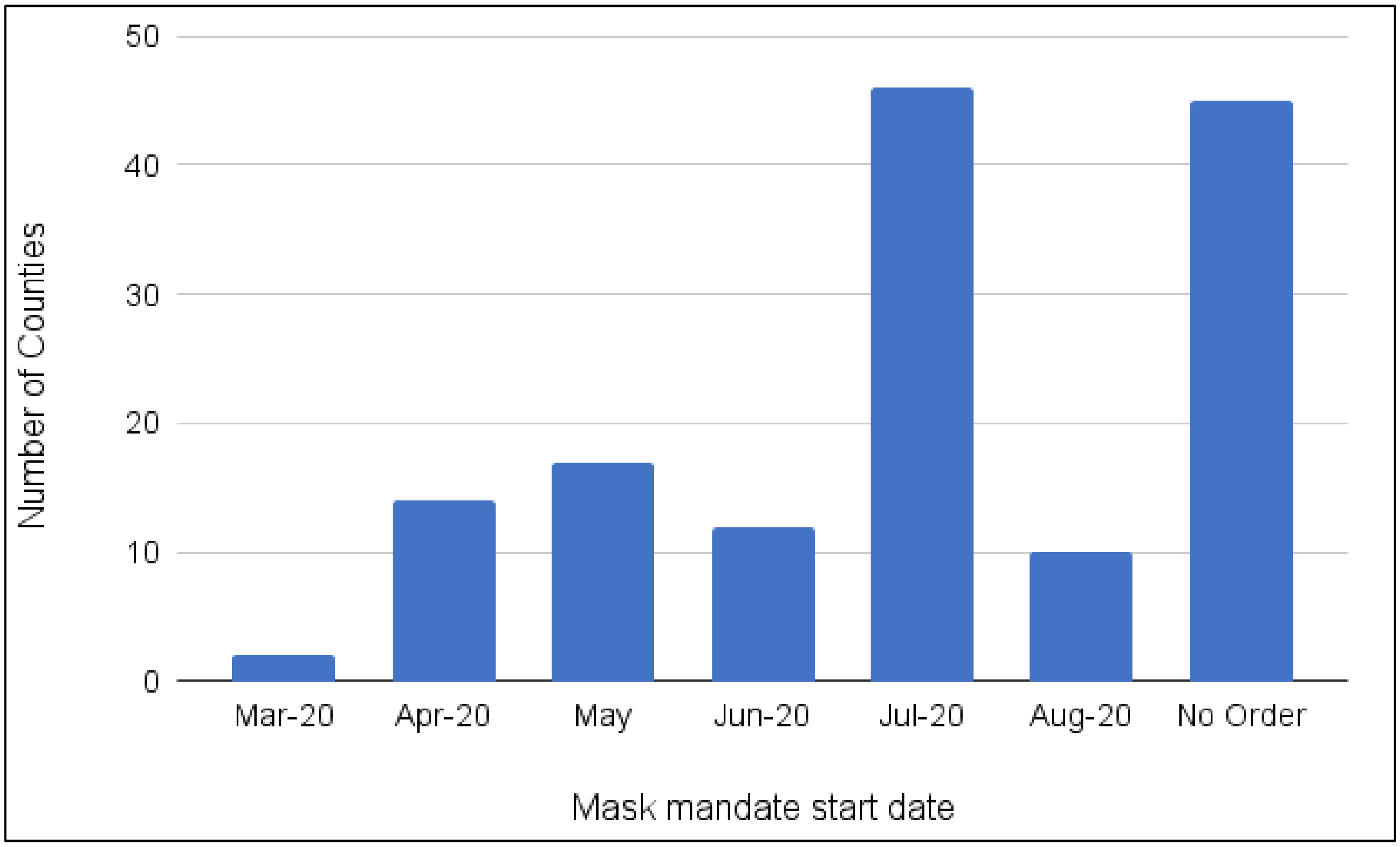

**Figure 7. Distribution of county-level mask mandate start dates**

## 4.6 Fusing Passive and Active Data

In this section, we show how the fusing of survey and POI data can help uncover underlying factors explaining behavioral change; we summarize the daily temporal coverage of survey respondent mobility data across all survey waves and illustrate how different groups can exhibit different mobility behavior. We define the daily data coverage by the number of panelists with POI data collected on each day. We cleaned the POI data to align the daily POI arrivals, trip starts, and POI departures across time for each survey respondent. In an ideal world, the POI data coverage would be 100% for all survey respondents. However, there were several instances where coverage was less than 100%. This could be due to several reasons, including but not limited to smartphone battery drain, GPS signal unavailability, or errors in the POI data collection process.

Not all study participants had mobility data available throughout the study duration. **Figure 8** shows the coverage of mobility data for survey respondents between January 1, 2020 and April 30, 2021. The share of panelists with mobility data available continuously increased from January 1, 2020 through its highest value of about 93% in early May 2020. Since then there was a decrease in mobility data availability to about 40% in May 2021.

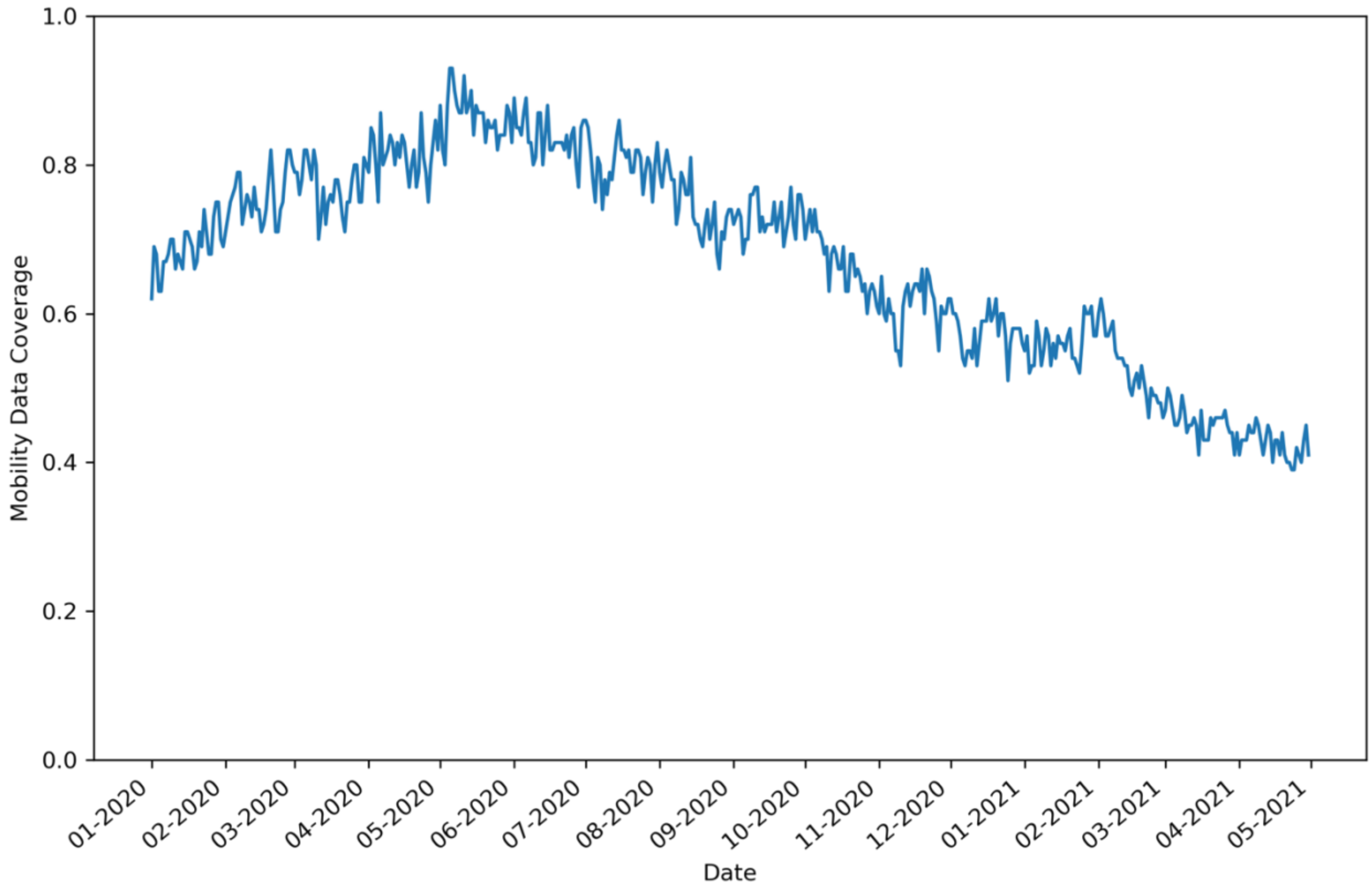

**Figure 8. Availability of mobility data from survey respondents**

We plotted the number of weekly trips made by first wave study participants between January 1, 2020 and April 30, 2021 in **Figure *9***. The figure confirms other research showing a sharp decrease in mobility in the early weeks of the pandemic, followed by a steady recovery since. We further explore changes in mobility behavior across several dimensions collected from the survey data in **Figure 10** through **Figure 12**. **Figure 10** shows that individuals with access to a car traveled significantly more than those without access, both before and after the beginning of the pandemic. Those in a households with vehicle access were able to reduce their travel significantly at the onset of the pandemic and their travel has recovered to pre-pandemic levels. Conversely, those without access to a car continued to exhibit lower mobility when compared to individuals with access to a vehicle. Similarly, **Figure 11** shows that study participants who believe that social distancing reduced COVID-19 spread exhibited lower mobility, compared to those who do not, in the early phases of the pandemic, although both groups show no statistically significant difference in mobility behavior at later stages of the pandemic. Additionally, individuals with low extraversion as measured by the BFI-10 questionnaire showed less mobility than those with average or low extraversion (**Figure 12**). Parker et al. (2021) used this POI data in conjunction with collected survey data to show how travel behavior has changed differently between public transit users and non-users due to the COVID-19 pandemic.

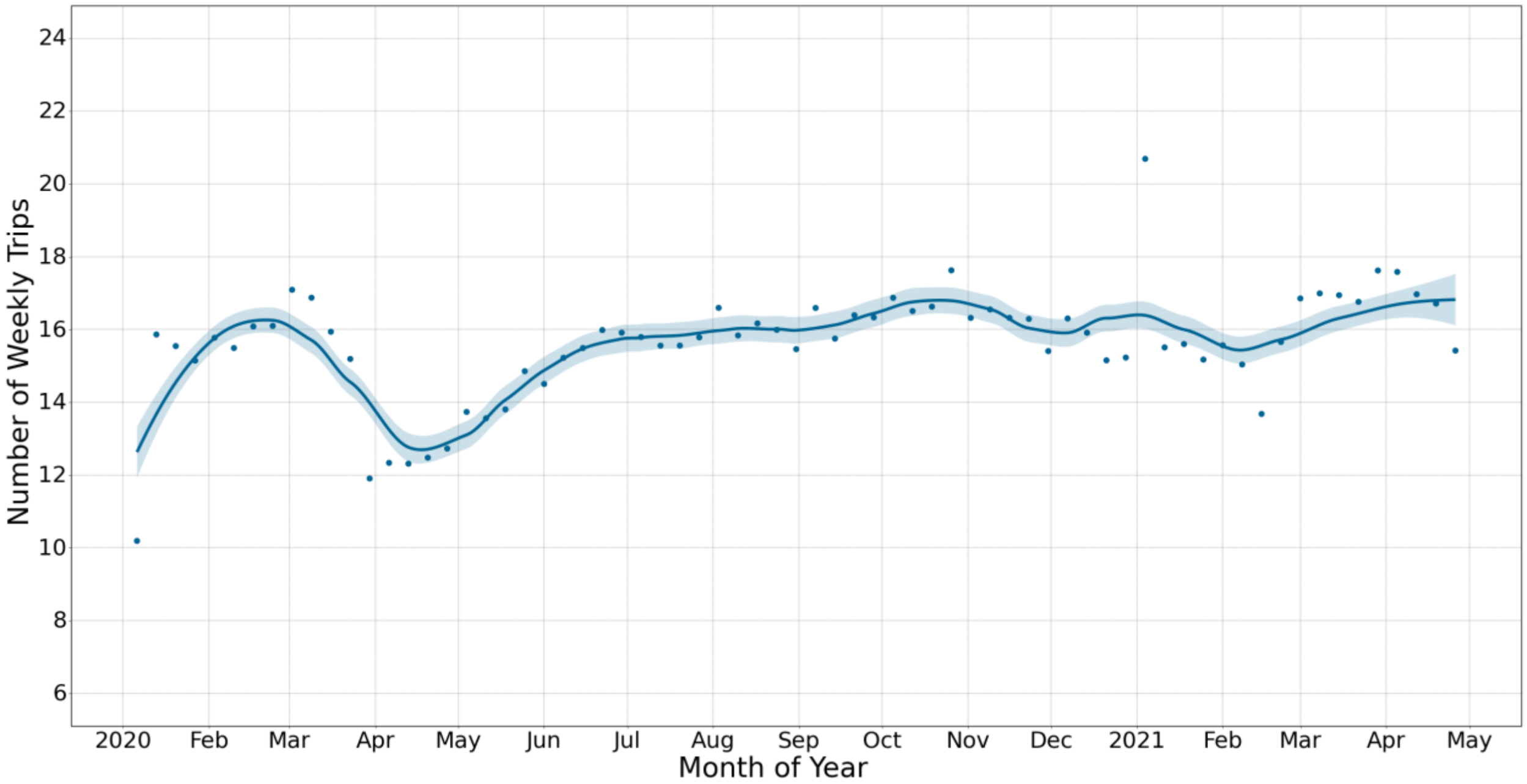

**Figure 9. Number of weekly trips by study participants**

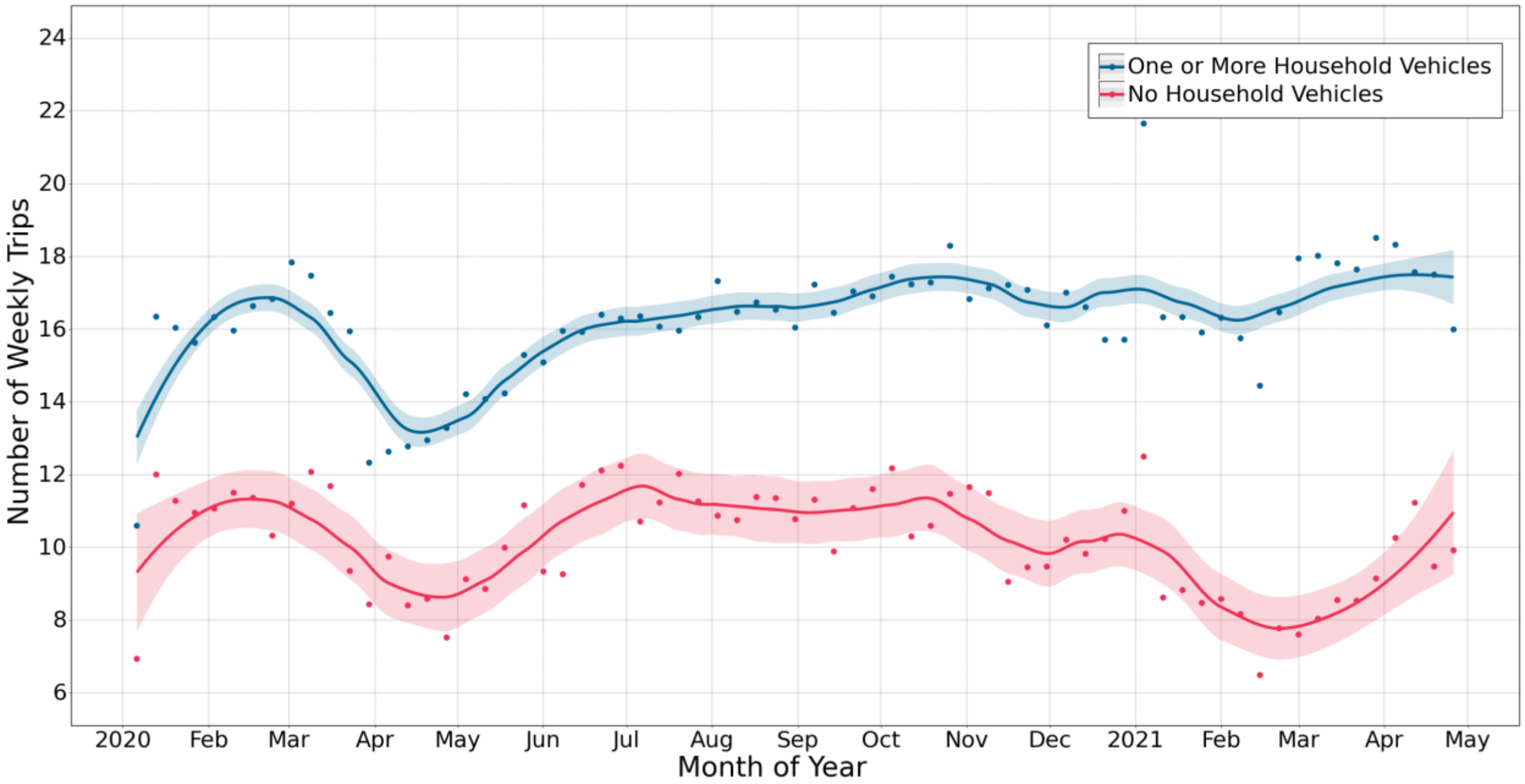

**Figure 10. Number of unique places visited weekly by study participants with and without access to a household vehicle**

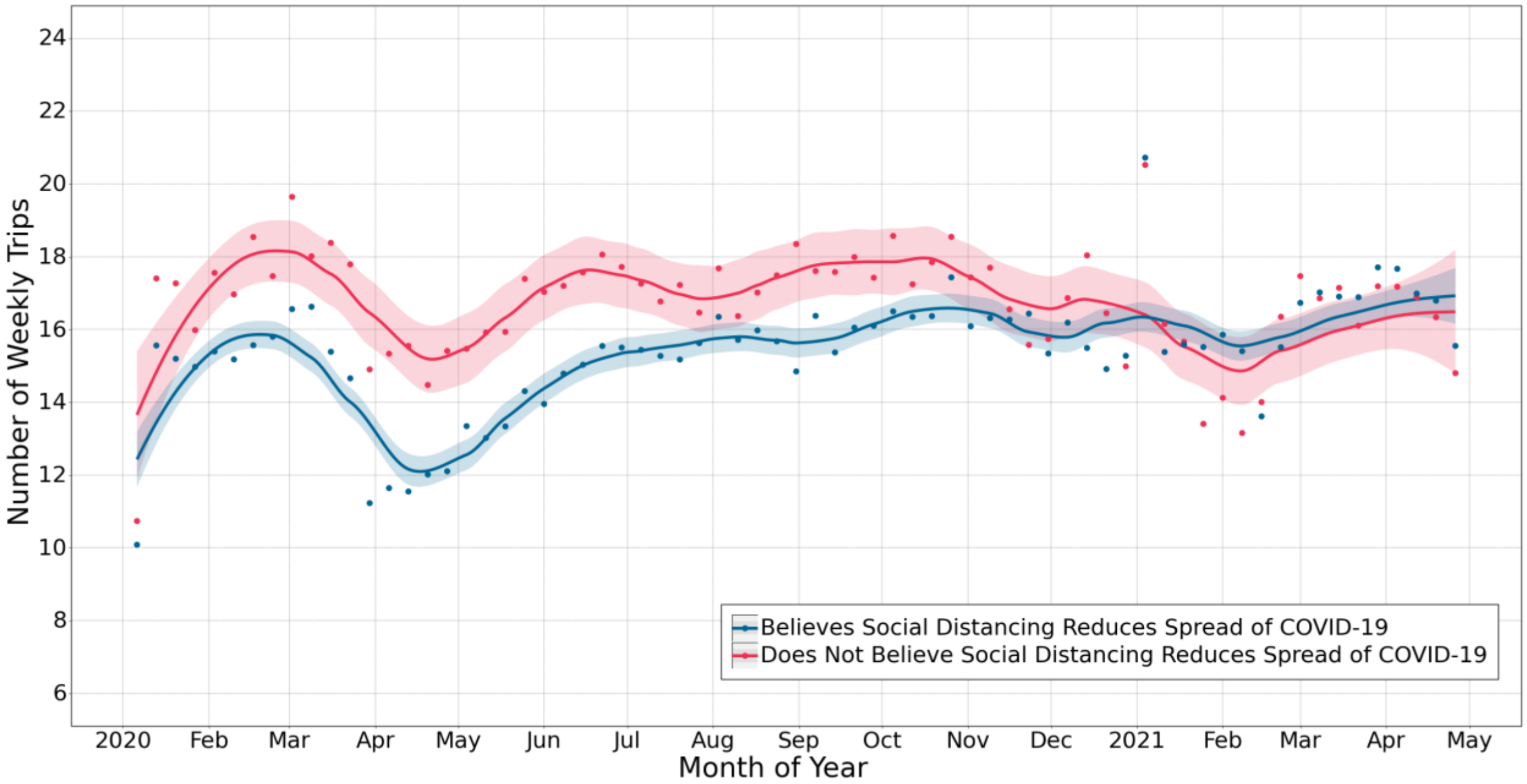

**Figure 11. Number of weekly trips by study participants who believe social distancing helps reduce COVID-19 spread vs. those who do not**

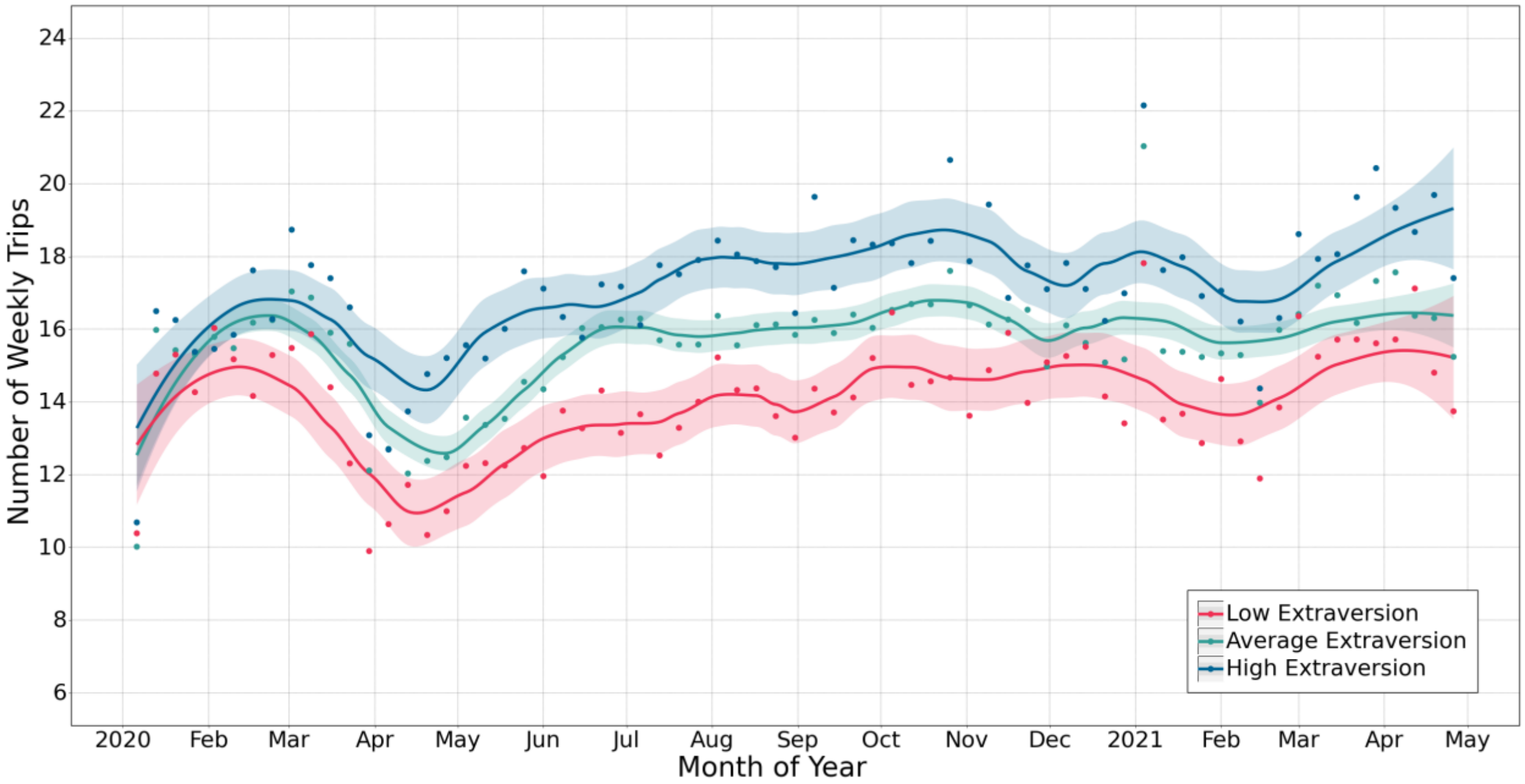

**Figure 12. Number of weekly trips by study participants with low, average, and high extraversion levels**

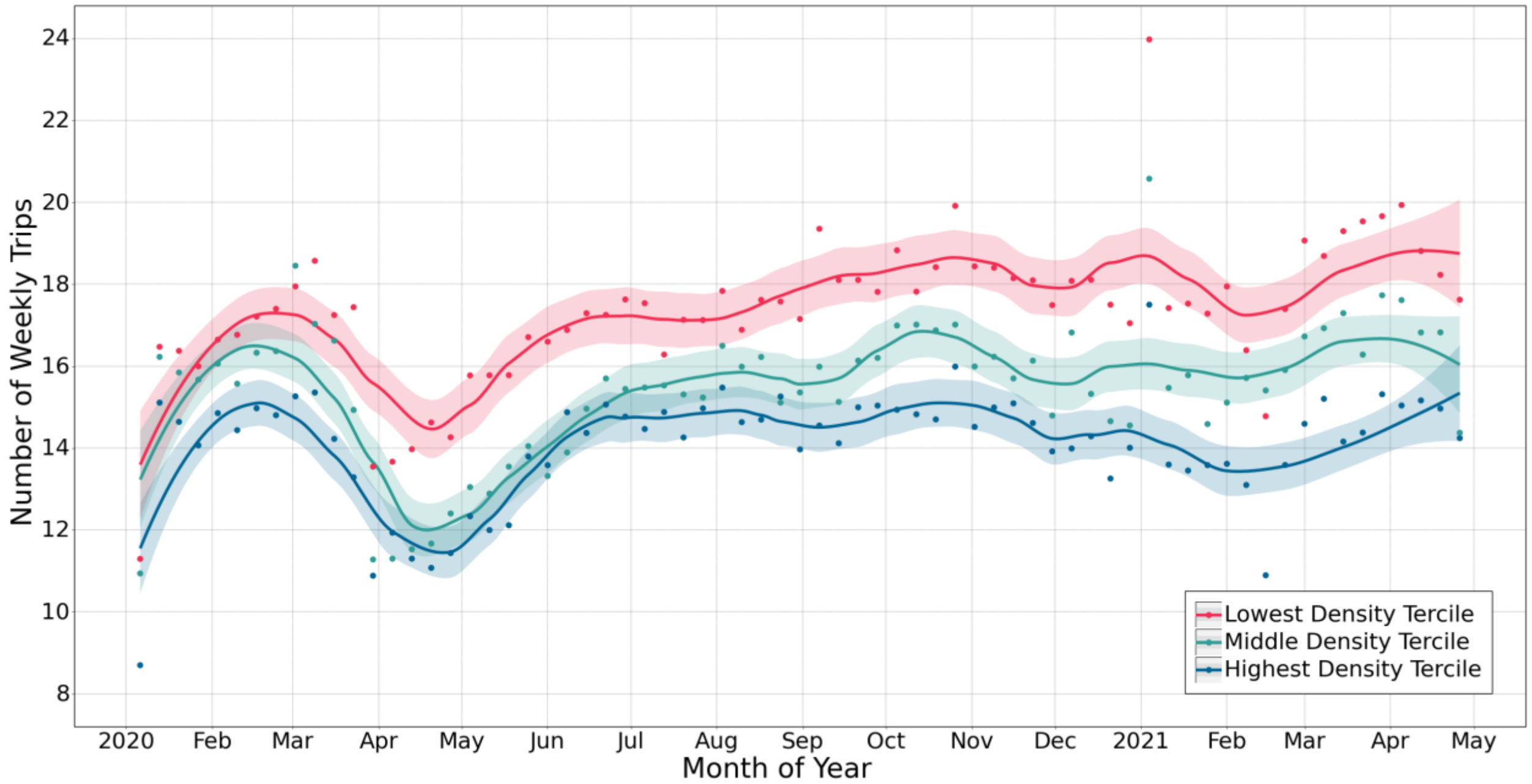

**Figure 14. Number of weekly trips by study participants from counties with the lowest, middle, and highest population density terciles**

# 5 Conclusions

In this paper, we presented a comprehensive summary of the data collection processes used for a set of COVID-19 studies. We collected longitudinal passive and active data from a set of U.S. panelists throughout the COVID-19 pandemic, starting in August 2020. The passive data consisted of individual level spatial POI data, which provides invaluable insight into human mobility behavior throughout the pandemic. From a review of the literature, and to the best of our knowledge, only one other study combines both individual passive (POI) and broad active (survey) data exist, particularly for a panel. The majority of other research uses only one of the two different sources. While there is some research that has collected both active and passive data, it has focused on a narrow set of research questions, in part because the active data are not comprehensive and because the data have not been collected on a longitudinal panel. Our longitudinal data reflects the different levels at which behavior can be influenced (individual vs. aggregate) and captures phenomena that vary at different time scales; we measure individual mobility on a daily basis, identifying several characteristics of daily travel behavior over time and capture COVID-19 related policies which are adjusted less frequently and are likely to have a broad impact on local or regional behavior.

This manuscript summarizes the study design and administration, briefly describes the collected data, and outlines some of the challenges faced throughout the study duration. This study collected data from five survey waves, starting in early August 2020, each of which lasted about a month. The surveys were slightly modified between waves to account for changes in the state of the pandemic in the U.S. All of the study participants were compensated for their participation in the study, with varying amounts in each of the survey waves to maintain the number of quality survey responses. Increases in the response rates through the survey waves suggest that the respondents remaining on the panel are more reliable panelists. Nonetheless, the research team still faced difficulties in achieving the desired response rates in later survey waves despite increasing survey incentives. This difficulty of getting the desired response rates could be due to several factors, including respondent survey fatigue. To overcome this hurdle, the research team added more panelists to refresh the respondents pool.

The descriptive statistics of this analysis show that our sample is not fully representative of the U.S. population, as it overrepresents low and medium income households, non-White Americans, and individuals with a college or university degree. Oversampling lower income households and non-White Americans presents a unique opportunity to garner insights on how the pandemic has specifically impacted these communities. Our sample is fairly representative of public transit and active mode users, but our longitudinal data does not reflect the drop in public transit use reported by other research in the early phases of the pandemic.

Our survey incompletion analysis also shows that our survey design might not have been properly suited for smartphones. A significant share of survey abandonment occurs at the mobility section of the survey. Several mobility questions were in a matrix format asking for several layers of information in a format not properly suited for smartphones. However, we notice that the overall survey incompletion rate has continuously decreased across the several survey waves, indicating that more panelists became more comfortable with the survey flow.

We also estimated several binary logit models to explore associations between sociodemographic characteristics and several aspects of survey participation, most notably survey participation, survey incompletion, failing the attention check questions, and returning to participate in the second survey wave. Results show that non-White respondents are more likely to not complete the survey and fail the attention checks. Additionally, panelists with a long history on the Similarweb platform are significantly less likely to participate in the study.

Our passive POI data confirms findings by other researchers, finding a significant drop in mobility activity in the early phases of the COVID-19 pandemic, followed by a consistent recovery since. We also show how fusing the passive POI data with active data could reveal mobility behavior heterogeneity across different groups; most notably, individuals without access to a car continued to exhibit significantly lower mobility when compared to individuals without one throughout the COVID-19 pandemic.

This study succeeded in collecting a comprehensive longitudinal panel dataset giving detailed insight into people's lives in response to the COVID-19 pandemic. The data collected is currently being used in multiple research efforts as outlined in the introduction and could be used in conjunction with other data sources to address a wider range of retrospective questions aiming at fully understanding the impacts of the COVID-19 pandemic and help shape our collective response to similar public health threats in the future.

# 6 Data Availability

The survey dataset will be made available to other researchers after the publication of current research based on the data, which we expect to be on the order of 2 years. We welcome collaboration; any researchers interested in using the dataset are encouraged to contact the authors. We aim to produce an open-source dataset based on the terms set by the U.C. Berkeley Institutional Review Board (identifying information cleaned) and the Similarweb IP that includes the complete survey data from the five waves and processed Similarweb passive data.

## Acknowledgements

The authors are grateful to Similarweb for providing access to the passive data used in this research, as well as access to their mobile platform to administer surveys to their panelists. The authors of this paper would like to thank the University of California Center for Information Technology Research in the Interest of Society (CITRIS) for funding support. Additionally, the authors would like to thank all the individuals who provided feedback throughout the duration of the study. The authors would also like to thank the participants for their valuable time. Any omissions are unintentional.

## Conflict of interest

The authors declare that they have no conflict of interest.

## Author Contributions

**Conceptualization:** Mohamed Amine Bouzaghrane, Hassan Obeid, Meiqing Li, Madeleine Parker, Daniel A. Rodríguez, Daniel G. Chatman, Karen Trapenberg Frick, Raja Sengupta, Joan Walker; **Methodology:** Mohamed Amine Bouzaghrane, Daniel A. Rodríguez, Daniel G. Chatman, Joan Walker; **Formal analysis and investigation:** Mohamed Amine Bouzaghrane, Hassan Obeid, Drake Hayes, Minnie Chen; **Writing - original draft preparation**: Mohamed Amine Bouzaghrane, Hassan Obeid, Minnie Chen, Drake Hayes, Meiqing Li, Madeleine Parker; **Writing - review and editing:** Daniel A. Rodríguez, Daniel G. Chatman, Karen Trapenberg Frick, Joan Walker; **Funding acquisition:** Daniel G. Chatman, Joan Walker ; **Supervision:** Joan Walker**.**

## Biography

**Mohamed Amine Bouzaghrane** is a Ph.D. student in the Civil and Environmental Engineering Department at UC Berkeley. His research interests are in the use of discrete choice modeling and causal inference to better inform transportation policies.

**Hassan Obeid** is a Ph.D. student in Civil & Environmental Engineering at the University of California, Berkeley. His research interests are in behavioral modeling, causal inference, and transportation systems.

**Drake Hayes** is a graduate student in the City and Regional Planning at UC. He is interested in researching what influences people's transportation mode choices and how to make transportation equitable, efficient, and environmentally friendly in North America.

**Minnie Chen** is an undergraduate student in the Civil and Environmental Engineering Department at UC Berkeley. Her research interests are in transportation modeling and travel behavior.

**Meiqing Li** is a PhD student in City and Regional Planning at UC Berkeley. Her research interests are sustainable mobility, built environment, and travel behavior modeling.

**Madeleine Parker** is a PhD student in City and Regional Planning at UC Berkeley. Her research interests are in climate change adaptation, resilience planning, affordable housing, transportation equity, urban mobility, spatial inequality, participatory mapping, data science, and spatial analysis.

**Daniel A. Rodríguez** is Chancellor's Professor of City and Regional Planning and Associate Director of the Institute for Transportation Studies. His research focuses on the relationship between transportation, land development, and the health and environmental consequences that follow.

**Daniel G. Chatman** is an Associate Professor in the Department of City and Regional Planning at the University of California, Berkeley. He studies travel behavior and the built environment; residential and workplace location choice; "smart growth" and municipal fiscal decision making; and the connections between public transportation, immigration and the economic growth of cities. His research relies heavily on original data collection, including surveys, focus groups and interviews.

**Karen Trapenberg Frick** is an Associate Professor in the Department of City and Regional Planning at the University of California, Berkeley. She also is Director of the University of California Transportation Center (UCTC). She is an expert on sustainable transport and community-based planning and major transportation infrastructure projects. Her current research focuses on conservative, Tea Party and property rights activists' perspectives on planning and planners' responses.

**Raja Sengupta** is currently Professor and Program Leader of the Systems program in Civil & Environmental Engineering at the University of California, Berkeley. He received his Ph.D from the EECS department of the University of Michigan, at Ann Arbor. His current research interests are in smart cities, mobile computing, cloud computing, human behavior change technology, transportation demand, robotics, drones, and control theory.

**Joan Walker** is a Professor in Civil & Environmental Engineering & Global Metropolitan Studies at the University of California, Berkeley. She received her Ph.D. from the Civil Engineering Department at MIT. Her current research interests are in behavioral modeling and transportation systems.

**Appendix A – Survey Questionnaire**

## Safety Measures

Q2.1 Thank you for choosing to take this survey! We'll be asking you a variety of questions related to your life during the Covid-19 pandemic.
Our first questions will help us learn about some of your recent habits.

Q2.2 In the **past seven days**, how many times a day have you been washing your hands with soap and water or using hand sanitizer?

- ○ 0 times
- ○ 1-3 times
- ○ 4-6 times
- ○ More than 6 times
- ○ I prefer not to answer

Q2.3 Relative to before the Covid-19 pandemic began (March 2020), how has your handwashing frequency changed?

- ○ Decreased
- ○ No change
- ○ Increased
- ○ I prefer not to answer

Q2.4 In the **past seven days**, how often have you worn a mask or face covering in the following situations?

| | Not applicable | Never | Sometimes | Frequently | Always | I prefer not to answer |
|---|---|---|---|---|---|---|
| **While inside your home** | ○ | ○ | ○ | ○ | ○ | ○ |
| **While indoors outside your home (e.g. grocery store, gym, etc.)** | ○ | ○ | ○ | ○ | ○ | ○ |
| **While walking or exercising on your own outside** | ○ | ○ | ○ | ○ | ○ | ○ |
| **While socializing with other people outside** | ○ | ○ | ○ | ○ | ○ | ○ |
| **While traveling in public (not in your own car)** | ○ | ○ | ○ | ○ | ○ | ○ |

Q2.5 Which type(s) of mask or face covering have you worn?

- ☐ I don't wear a mask
- ☐ Cloth mask
- ☐ Surgical mask
- ☐ N95 or similar
- ☐ Scarf, bandana or other cloth face covering
- ☐ A mesh mask that intentionally doesn’t provide protection
- ☐ Other ________________________________________________
- ☐ I prefer not to answer

Q2.6 In the **last two weeks**, what is the largest group gathering you have been to outside of your household?

- ○ I haven't been to a group gathering outside my household
- ○ Fewer than 5 people
- ○ Between 5 and 10 people
- ○ Between 10 and 20 people
- ○ Between 20 and 50 people
- ○ Between 50 and 100 people
- ○ More than 100 people
- ○ I don't know
- ○ I prefer not to answer

Q2.7 In the **last two weeks**, which of the following actions did you take because of the Covid-19 pandemic?

- ☐ Cancelled or changed travel plans
- ☐ Did not go to religious or other community events
- ☐ Held no in-person gatherings with friends
- ☐ Sanitized groceries or other purchases
- ☐ Stayed at least 6ft away from other people in public
- ☐ Avoided shaking hands or hugging others not in my household
- ☐ Other ______________________________________________
- ☐ None of the above
- ☐ I prefer not to answer

## Mobility

Q3.1 These next few questions are about how you travel on a daily basis.

Q3.2 Have you been employed at all in the **last 12 months?**

- ○ Yes
- ○ No
- ○ I prefer not to answer

Q3.3 Please indicate how many days you worked in the **last seven days**

- ○ 0
- ○ 1
- ○ 2
- ○ 3
- ○ 4
- ○ 5
- ○ 6
- ○ 7
- ○ I prefer not to answer

Q3.4 Please indicate how many days you worked **from home** in the **last seven days**

- ○ 0
- ○ 1
- ○ 2
- ○ 3
- ○ 4
- ○ 5
- ○ 6
- ○ 7
- ○ I prefer not to answer

Q3.5 Relative to before the Covid-19 pandemic began (March 2020), how has the number of days you travel to work changed?

- ○ Significantly decreased
- ○ Somewhat decreased
- ○ No change
- ○ Somewhat increased
- ○ Significantly increased
- ○ I prefer not to answer

Q3.6 If you worked in the **last seven days**, which was the primary means of transportation you used to get to work?

- ○ I did not leave my home for work
- ○ Driving alone or with household members
- ○ Carpooling with people outside my household
- ○ Carsharing (e.g. Zipcar, Gig, etc.)
- ○ Ridehail (e.g. Uber/Lyft) or Taxi
- ○ Transit (e.g. bus, subway, train, etc.)
- ○ Bicycle
- ○ Walking
- ○ Other ________________________________________________
- ○ I prefer not to answer

Q3.7 In the **last seven days**, on how many days did you use the following for transportation (not including walks around your neighborhood for exercise, etc.)?

| | 0 days | 1-2 days | 3 or more days | I prefer not to answer |
|---|---|---|---|---|
| **Driving alone or with household members** | ○ | ○ | ○ | ○ |
| **Carpooling with people outside my household** | ○ | ○ | ○ | ○ |
| **Carsharing (e.g. Zipcar, Gig, etc.)** | ○ | ○ | ○ | ○ |
| **Ridehail (e.g. Uber/Lyft) or Taxi** | ○ | ○ | ○ | ○ |
| **Transit (e.g. bus, subway, train, etc.)** | ○ | ○ | ○ | ○ |
| **Bicycle** | ○ | ○ | ○ | ○ |
| **Walking** | ○ | ○ | ○ | ○ |

Q3.8 Is this more, less, or about the same as your use of each before Covid-19?

| | Less | About the same | More | I prefer not to answer |
|---|---|---|---|---|
| **Driving alone or with household members** | ○ | ○ | ○ | ○ |
| **Carpooling with people outside my household** | ○ | ○ | ○ | ○ |
| **Carsharing (e.g. Zipcar, Gig, etc.)** | ○ | ○ | ○ | ○ |
| **Ridehail (e.g. Uber/Lyft) or Taxi** | ○ | ○ | ○ | ○ |
| **Transit (e.g. bus, subway, train, etc.)** | ○ | ○ | ○ | ○ |
| **Bicycle** | ○ | ○ | ○ | ○ |
| **Walking** | ○ | ○ | ○ | ○ |

Q3.9 In the **last seven days**, for which purposes did you use each of the following for transportation?

| | Work/school | Healthcare | Errands (e.g. groceries) | Leisure | Other | I prefer not to answer |
|---|---|---|---|---|---|---|
| **Driving alone or with household members** | ☐ | ☐ | ☐ | ☐ | ☐ | ☐ |
| **Carpooling with people outside my household** | ☐ | ☐ | ☐ | ☐ | ☐ | ☐ |
| **Carsharing (e.g. Zipcar, Gig, etc.)** | ☐ | ☐ | ☐ | ☐ | ☐ | ☐ |
| **Ridehail (e.g. Uber/Lyft) or Taxi** | ☐ | ☐ | ☐ | ☐ | ☐ | ☐ |
| **Transit (e.g. bus, subway, train, etc.)** | ☐ | ☐ | ☐ | ☐ | ☐ | ☐ |
| **Bicycle** | ☐ | ☐ | ☐ | ☐ | ☐ | ☐ |
| **Walking** | ☐ | ☐ | ☐ | ☐ | ☐ | ☐ |

Q3.10 Have cuts in transit service since the Covid-19 pandemic began (March 2020) been an issue for you?

- ○ Yes, a significant issue
- ○ Yes, a minor issue
- ○ No, they haven't been an issue
- ○ There have not been any service cuts
- ○ I do not use transit
- ○ I prefer not to answer

Q3.11 Which of the following measures would increase your use of transit (including bus, subway, train, etc.)?

- ☐ Widespread use of face masks
- ☐ Reduced crowding
- ☐ Reduction of Covid-19 rates in area
- ☐ Effective Covid-19 treatment or vaccine
- ☐ Increased sanitation/cleaning
- ☐ Return to regular service levels/schedule frequency
- ☐ I am already comfortable using transit
- ☐ None of the above
- ☐ Other: ________________________________________________
- ☐ I prefer not to answer

Q3.12 Which of the following measures would increase your use of ride hailing or sharing services such as Uber, Lyft, Zipcar or Gig?

- ☐ Widespread use of face masks
- ☐ Reduction of Covid-19 rates in area
- ☐ Effective Covid-19 treatment or vaccine
- ☐ Increased sanitation/cleaning
- ☐ I am already comfortable using these modes
- ☐ None of the above
- ☐ Other: ________________________________________________
- ☐ I prefer not to answer

Q3.13 If you are reading this question, please select "Somewhat increased" below

- ○ Significantly decreased
- ○ Somewhat decreased
- ○ No change
- ○ Somewhat increased
- ○ Significantly increased
- ○ I prefer not to answer

Q3.14 How many vehicles are available for regular use by the people who currently live in your household?

- ○ 0
- ○ 1
- ○ 2
- ○ 3+
- ○ I prefer not to answer

Q3.15 As a result of the Covid-19 pandemic, have you made any of the following purchases? Please only select those that you wouldn't have made if it weren't for the pandemic.

- ☐ Purchases to improve your home/outdoor living space
- ☐ Purchases to improve your home working/school environment
- ☐ Purchases to improve your home or personal security
- ☐ Purchases to support your physical health
- ☐ Purchases to support a hobby
- ☐ Purchases to increase your motorized transportation options (car, RVs, etc.)
- ☐ Purchases to increase your active transportation options (bicycle, etc.)
- ☐ Other ________________________________________________
- ☐ None
- ☐ I prefer not to answer

## Household Dynamics

Q4.1 These next questions are about who you currently live with.

Q4.2 How many people (adults and children) currently live in your household, including yourself?

- ○ 1 (I live alone)
- ○ 2
- ○ 3
- ○ 4
- ○ 5
- ○ 6+
- ○ I prefer not to answer

Q4.3 How does your relationship with your household members now compare to before Covid-19?

- ○ Much worse
- ○ Somewhat worse
- ○ About the same
- ○ Somewhat better
- ○ Much better
- ○ I don't know
- ○ I prefer not to answer

Q4.4 Does conflict within your household affect your ability to spend longer periods of time at home?

- ○ Yes
- ○ Somewhat
- ○ No
- ○ I don't know
- ○ I prefer not to answer

## Economic Factors

Q5.1 These next questions are about how your employment is changing with Covid-19.

Q5.2 Were you working before the Covid-19 pandemic started?

- ○ Yes, full-time
- ○ Yes, part-time
- ○ No
- ○ I prefer not to answer

Q5.3 Since the beginning of the Covid-19 pandemic (March 2020), have you experienced any of the following changes to your working situation?

- ☐ Been laid off or lost a job (Approximate date: mm/dd/yyyy)
- ☐ Reduced pay or income
- ☐ Put on temporary leave from job
- ☐ Increased hours worked per week
- ☐ Got a new job
- ☐ Job did not change
- ☐ Other
- ☐ I prefer not to answer

Q5.4 Since the beginning of the Covid-19 pandemic (March 2020), have you received any form of financial government assistance?

- ○ Yes
- ○ No
- ○ I prefer not to answer

Q5.5 Do you believe your change in employment was a result of the Covid-19 pandemic?

- ○ Yes
- ○ No
- ○ I don't know
- ○ I prefer not to answer

Q5.6 Relative to before the Covid-19 pandemic began (March 2020), how has your household income changed?

- ○ Significantly decreased
- ○ Somewhat decreased
- ○ No change
- ○ Somewhat increased
- ○ Significantly increased
- ○ I prefer not to answer

Q5.7 How confident are you that your household will be able to pay your next rent or mortgage payment on time?

- ○ Not confident
- ○ Somewhat confident
- ○ Very confident
- ○ Payment is/will be deferred
- ○ Not applicable
- ○ I prefer not to answer

Q5.8 How would a $400 emergency expense that you may have to pay impact your ability to pay your other bills this month?

- ○ I would not be able to pay all of my bills
- ○ I could still pay all of my bills
- ○ I don’t know
- ○ I prefer not to answer

Q5.9 How much longer do you think you can endure the **economic** impact of Covid-19?

- I can't endure it anymore
- A few more weeks
- A few more months
- A year
- Indefinitely
- I don't know
- I prefer not to answer

Q5.10 How much longer do you think you can endure the **emotional** impact of Covid-19?

- I can't endure it anymore
- A few more weeks
- A few more months
- A year
- Indefinitely
- I don't know
- I prefer not to answer

# Political

Q6.1 The next questions relate to your views and activities.

Q6.2 How closely do you follow the news?

- ○ Not at all closely
- ○ Not too closely
- ○ Fairly closely
- ○ Very closely
- ○ I prefer not to answer

Q6.3 Which of the following are your **main** sources of Covid-19 news?

- ☐ National cable news outlets, e.g. Fox News, CNN, MSNBC
- ☐ Nationally broadcast television news, e.g. ABC, CBS, NBC, PBS
- ☐ Local news outlets
- ☐ Radio and/or podcasts
- ☐ National newspapers (printed or online)
- ☐ Social Media, e.g. Facebook, Twitter, Instagram
- ☐ State/local officials
- ☐ Public health organizations and officials
- ☐ The White House & the Coronavirus Task Force
- ☐ Family and friends
- ☐ Other ________________________________________________
- ☐ I prefer not to answer

Q6.4 Which of the following Covid-19 related guidelines are currently in effect in your local area?

- ☐ No indoor dining
- ☐ No outdoor dining
- ☐ Some businesses are not allowed to open (e.g. gyms, salons, bars, movie theaters, etc.)
- ☐ Face coverings required indoors outside your home (e.g. grocery store)
- ☐ Face coverings required on public transportation (e.g. bus, subway, train, etc.)
- ☐ Face coverings required when outdoors in public spaces
- ☐ Restrictions on in-person gatherings (e.g. community, religious, entertainment, social, etc.)
- ☐ Quarantine and/or testing required for out-of-state visitors
- ☐ Restrictions on residential evictions
- ☐ None
- ☐ Other ________________________________________________
- ☐ I don't know
- ☐ I prefer not to answer

Q6.5 What do you think about Covid-19 related restrictions imposed on your local area?

- ○ Far too lenient
- ○ Too lenient
- ○ About right
- ○ Too strict
- ○ Far too strict
- ○ I don't know
- ○ I prefer not to answer

Q6.6 What do most of your family members and friends think about social distancing and stay-at-home directives imposed on your local area?

- ○ Far too lenient
- ○ Too lenient
- ○ About right
- ○ Too strict
- ○ Far too strict
- ○ I don't know
- ○ I prefer not to answer

Q6.7 Do you feel people in your local area are complying with the imposed Covid-19 related restrictions?

- ○ No, almost all are not complying
- ○ No, many are not complying
- ○ Yes, many are complying
- ○ Yes, almost all are complying
- ○ I don't know
- ○ Not applicable
- ○ I prefer not to answer

Q6.8 Since the COVID-19 pandemic began (March 2020), which of these activities did you leave your house for?

- ☐ Eating outdoors at a restaurant/bar
- ☐ Eating indoors at a restaurant/bar
- ☐ Grocery shopping
- ☐ Attending a large sports or concert event
- ☐ Attending a protest
- ☐ Going to the gym
- ☐ Medical/healthcare
- ☐ Caring for a relative or friend
- ☐ Socializing with friends in person
- ☐ Attending religious services
- ☐ Flying on an airplane
- ☐ Staying at a hotel/airbnb
- ☐ Other ________________________________________________
- ☐ I prefer not to answer

Q6.9 Which of the following form(s) of protests have you attended since the Covid-19 pandemic began?

☐ Against police brutality/systemic racism

☐ Against current business/work/school closures and in support of reopening areas

☐ Against vaccines

☐ Other ______________________________________________

☐ I prefer not to answer

1 Q6.10 What do you think about the following statements?

| | Strongly disagree | Disagree | Agree | Strongly agree | I prefer not to answer |
|---|---|---|---|---|---|
| Small businesses (e.g., local restaurants and bars) **could not** survive if people keep social distancing | ○ | ○ | ○ | ○ | ○ |
| Young adults **do not** need to practice social distancing | ○ | ○ | ○ | ○ | ○ |
| Social distancing stops Covid-19 from spreading around | ○ | ○ | ○ | ○ | ○ |
| Older adults should stay at home because they are more vulnerable | ○ | ○ | ○ | ○ | ○ |
| Not being able to hang out makes me upset | ○ | ○ | ○ | ○ | ○ |
| Social distancing decreases the burden on medical resources, so people in need can use them | ○ | ○ | ○ | ○ | ○ |
| Social distancing makes people lose their jobs | ○ | ○ | ○ | ○ | ○ |
| The government **should not** mandate wearing masks/face coverings | ○ | ○ | ○ | ○ | ○ |
| Wearing masks reduces the spread of Covid-19 | ○ | ○ | ○ | ○ | ○ |
| I practice social distancing because people around me do so | ○ | ○ | ○ | ○ | ○ |
| Federal government assistance, **in the form of stimulus checks to individuals**, is the right thing to do | ○ | ○ | ○ | ○ | ○ |
| Federal government assistance, **in the form of aid to businesses**, is the right thing to do | ○ | ○ | ○ | ○ | ○ |

1

Q6.11 Are you registered to vote?

- ○ Yes
- ○ No
- ○ I don't know
- ○ Not eligible
- ○ I prefer not to answer

Q6.12 How often have you voted since you became eligible?

- ○ Every election
- ○ Almost every election
- ○ Some elections
- ○ Rarely
- ○ There hasn't been an election since I became eligible
- ○ I don’t vote
- ○ Not Eligible
- ○ I prefer not to answer

Q6.13 Generally speaking, would you describe your political views as …?

- ○ Very conservative
- ○ Somewhat conservative
- ○ Moderate
- ○ Somewhat liberal
- ○ Very liberal
- ○ I prefer not to answer

Q6.14 How important is your religious community in your life?

- ○ Not at all important
- ○ Not too important
- ○ Fairly important
- ○ Very important
- ○ Not applicable
- ○ I prefer not to answer

Q6.15 Would you get a Covid-19 vaccine if/when one becomes available?

- ○ Definitely not
- ○ Probably not
- ○ Probably
- ○ Definitely
- ○ I don't know
- ○ I prefer not to answer

Q6.16 What would be a reason(s) not to get a Covid-19 vaccine?

- ☐ I would be concerned about side effects or getting infected from the vaccine
- ☐ I’m not concerned about getting seriously ill from Covid-19
- ☐ I don’t think vaccines work very well
- ☐ I won’t have time to get vaccinated
- ☐ Other ________________________________________________
- ☐ I prefer not to answer

# Personality

Q7.1 The next few questions ask about how you think of yourself.

Q7.2 How well do the following statements describe your personality?

I see myself as someone who...

| | Strongly disagree | Disagree | Neither agree nor disagree | Agree | Strongly agree | I prefer not to answer |
|---|---|---|---|---|---|---|
| **Is reserved** | ○ | ○ | ○ | ○ | ○ | ○ |
| **Is generally trusting** | ○ | ○ | ○ | ○ | ○ | ○ |
| **Tends to be lazy** | ○ | ○ | ○ | ○ | ○ | ○ |
| **Is relaxed, handles stress well** | ○ | ○ | ○ | ○ | ○ | ○ |
| **Has few artistic interests** | ○ | ○ | ○ | ○ | ○ | ○ |
| **Is outgoing, sociable** | ○ | ○ | ○ | ○ | ○ | ○ |
| **Tends to find fault with others** | ○ | ○ | ○ | ○ | ○ | ○ |
| **Does a thorough job** | ○ | ○ | ○ | ○ | ○ | ○ |
| **Gets nervous easily** | ○ | ○ | ○ | ○ | ○ | ○ |
| **Has an active imagination** | ○ | ○ | ○ | ○ | ○ | ○ |
| **If you're reading this, please select Disagree** | ○ | ○ | ○ | ○ | ○ | ○ |
| **Is willing to take risks** | ○ | ○ | ○ | ○ | ○ | ○ |
| **Is influenced by people I am close to** | ○ | ○ | ○ | ○ | ○ | ○ |

## Physical Health

Q8.1 Thanks for sticking with us! These next questions ask about your health.

Q8.2 How would you rate your physical health?

- ○ Poor
- ○ Fair
- ○ Good
- ○ Very good
- ○ Excellent
- ○ I Don't know
- ○ I prefer not to answer

Q8.3 Which of the following is your **main** source of healthcare coverage?

- ○ Employer sponsored (yourself, spouse, or parents)
- ○ Personally purchased
- ○ Student health care plan
- ○ Medicare/Medicaid/Other government healthcare plan
- ○ Do not have insurance
- ○ Other Source ________________________________________________
- ○ I don't know
- ○ I prefer not to answer

Q8.4 Does anyone in your household **not** have health insurance or some other kind of healthcare plan?

- ○ Yes
- ○ No
- ○ I don't know
- ○ I prefer not to answer

Q8.5 Do you suspect you have ever been infected with Covid-19?

- ○ Yes
- ○ No
- ○ I don't know
- ○ I prefer not to answer

Q8.6 Have you been tested for Covid-19 or the antibodies to it?

- ○ No, not tested
- ○ Yes, tested positive
- ○ Yes, tested negative
- ○ Yes, prefer not to specify result
- ○ I prefer not to answer

Q8.7 Do you think someone in your household has had Covid-19?

- ○ Yes
- ○ No
- ○ I don't know
- ○ I prefer not to answer

Q8.8 Do you personally know someone else outside of your household (relative, coworker, friend, etc.) who has had Covid-19?

- Yes
- No
- I don't know
- I prefer not to answer

Q8.9 Have you been hospitalized due to Covid-19, or do you personally know someone who has been hospitalized or died due to Covid-19?

- ○ Yes
- ○ No
- ○ I don't know
- ○ I prefer not to answer

Q8.10 In the **last seven days**, did you experience the following symptoms?

| | Yes | No | I prefer not to answer |
|---|---|---|---|
| **Severe or significant persistent cough** | ○ | ○ | ○ |
| **Shortness of breath** | ○ | ○ | ○ |
| **Fever higher than 100°F** | ○ | ○ | ○ |
| **New loss of smell and/or taste** | ○ | ○ | ○ |
| **Sore throat** | ○ | ○ | ○ |
| **Severe fatigue** | ○ | ○ | ○ |
| **Skipped meals** | ○ | ○ | ○ |

Q8.11 How worried are you that you or someone in your family will get Covid-19?

- ○ Very worried
- ○ Somewhat worried
- ○ Not too worried
- ○ Not at all worried
- ○ I/they already have
- ○ I don't know
- ○ I prefer not to answer

## Psychological Factors

Q9.1 Here are a few more questions that ask about how you're doing.

Q9.2 Relative to before the Covid-19 pandemic began (March 2020), how has your ability to get things done for work changed?

- ○ Significantly decreased
- ○ Somewhat decreased
- ○ No change
- ○ Somewhat increased
- ○ Significantly increased
- ○ Not applicable
- ○ I prefer not to answer

Q9.3 Since the Covid-19 pandemic began (March 2020), how often do you feel isolated or lacking companionship?

- ○ Never
- ○ Rarely
- ○ Sometimes
- ○ Often
- ○ Always
- ○ I prefer not to answer

Q9.4 Is this more, less, or about the same as before Covid-19?

- ○ Less
- ○ About the same
- ○ More
- ○ I prefer not to answer

Q9.5 Over the **last two weeks**, how often have you been bothered by the following problems?

| | Not at all | Several days | More than half the days | Nearly every day | I prefer not to answer |
|---|---|---|---|---|---|
| Feeling nervous, anxious, or on edge | ○ | ○ | ○ | ○ | ○ |
| Not being able to stop or control worrying | ○ | ○ | ○ | ○ | ○ |
| Little interest or pleasure in doing things | ○ | ○ | ○ | ○ | ○ |
| Feeling down, depressed, or | ○ | ○ | ○ | ○ | ○ |

hopeless

## Demographics

Q10.1 Finally, these last few questions ask about your current living situation.

Q10.2 What type of building do you currently live in?

- A mobile home
- A one-family house detached from any other house
- A one-family house attached to one or more houses
- A building with 2 to 5 apartments
- A building with 6 to 19 apartments
- A building with 20 or more apartments
- Other (boat, RV, van, tent, etc.)
- I prefer not to answer

Q10.3 Which of the following best describes your place of residence?

- Owned free and clear
- Owned with a mortgage or loan
- Rented
- Occupied without payment of rent
- I don't have a stable place of residence
- I prefer not to answer

Q10.4 How many children ages 0-6 currently live in your household?

- 0
- 1
- 2
- 3+
- I prefer not to answer

Q10.5 How many children ages 7-17 currently live in your household?

- 0
- 1
- 2
- 3+
- I prefer not to answer

Q10.6 Do you currently help care for an elderly or disabled person (family, relative, or friend)?

- Yes, at my home
- Yes, outside my home
- No
- I prefer not to answer

Q10.7 Do you currently help care for a child (family, relative, or friend)?

- Yes, at my home
- Yes, outside my home
- No
- I prefer not to answer

Q10.8 Are you an “essential worker”?

- ☐ No
- ☐ Yes, a healthcare worker
- ☐ Yes, a grocery store, pharmacy, or retail worker
- ☐ Yes, a food service worker
- ☐ Yes, a child care worker
- ☐ Yes, an essential government worker (e.g. firefighter, police officer)
- ☐ Other ________________________________________________
- ☐ I prefer not to answer

Q10.9 Is anyone else in your household (excluding yourself) classified as an “essential worker”?

- ☐ No
- ☐ Yes, a healthcare worker
- ☐ Yes, a grocery store, pharmacy, or retail worker
- ☐ Yes, a food service worker
- ☐ Yes, a child care worker
- ☐ Yes, an essential government worker (e.g. firefighter, police officer)
- ☐ Other ________________________________________________
- ☐ I prefer not to answer

Q10.10 How many **bedrooms** are in your home?

- ○ 0
- ○ 1
- ○ 2
- ○ 3
- ○ 4
- ○ 5
- ○ 6+
- ○ I prefer not to answer

Q10.11 How many **bathrooms** are in your home?

- ○ 0
- ○ 1
- ○ 2
- ○ 3
- ○ 4
- ○ 5
- ○ 6+
- ○ I prefer not to answer

Q10.12 How many **other rooms** are in your home? (exclude hallways, balconies, foyers and porches)

○ 0

○ 1

○ 2

○ 3

○ 4

○ 5

○ 6+

○ I prefer not to answer

○

## Open Ended Questions

Q11.1 Thank you for taking this survey! The final three questions give you an opportunity to tell us more about your experiences if you wish.

Q11.2 Is there anything else you would like to tell us about things that are particularly difficult for you during the pandemic?

________________________________________________________________

________________________________________________________________

Q11.3 Is there anything else you would like to tell us about positive aspects of your life during the pandemic?

________________________________________________________________

________________________________________________________________

Q11.4 Is there anything you would like to tell us about our survey?

________________________________________________________________

________________________________________________________________

## Appendix B – Survey Burden According to Axhausen et al. (2015)

| Survey Section | Wave 1 (Aug.2020) | Wave 2 (Oct.2020) | Wave 3 (Dec.2020) | Wave 4 (Apr.2021) | Wave 5 (Jul.2021) |
|---|---|---|---|---|---|
| **Safety Measures** | 58 | 62 | 46 | 74 | 42 |
| **Modality** | 195 | 195 | 157 | 190 | 157 |
| **Household Dynamics** | 15 | 41 | 41 | 41 | 41 |
| **Economic Factors** | 50 | 49 | 46 | 49 | 46 |
| **Political Factors** | 162 | 159 | 156 | 167 | 189 |
| **Personality** | 52 | 52 | N/A | 52 | N/A |
| **Physical Health** | 70 | 70 | 70 | 70 | 70 |
| **Psychological Factors** | 30 | 30 | 26 | 30 | 26 |
| **Demographics** | 67 | 71 | 59 | 84 | 69 |
| **Open Ended Questions** | 18 | 18 | 18 | 18 | 18 |